\documentclass[reprint,aps,showkeys,superscriptaddress,
 prx,amsmath,amssymb,bm
]{revtex4-2}
\usepackage{graphicx}
\usepackage{subfiles}
\usepackage{hyperref}
\hypersetup{
    colorlinks=true,
    linkcolor=blue,
    citecolor=blue,
    urlcolor=blue,
    pdftitle={IC-ZNE}
    }

\begin{document}

\preprint{APS/123-QED}

\title{Inverted-circuit zero-noise extrapolation for quantum gate error mitigation}

\author{Kathrin F. Koenig}
\email{kathrin.koenig(at)iaf.fraunhofer.de}
\affiliation{
 Fraunhofer Institute for Applied Solid State Physics, Tullastr. 72, 79108 Freiburg, Germany
}
\affiliation{Department of Sustainable Systems Engineering, University of Freiburg, Emmy-Noether-Str. 2, 79110 Freiburg, Germany}

\author{Finn Reinecke}
\affiliation{
 Fraunhofer Institute for Applied Solid State Physics, Tullastr. 72, 79108 Freiburg, Germany
}
\affiliation{Department of Computer Science, University of Freiburg, Georges-K{\"o}hler-Allee 51,
79110 Freiburg, Germany}

\author{Walter Hahn}
\affiliation{
 Fraunhofer Institute for Applied Solid State Physics, Tullastr. 72, 79108 Freiburg, Germany
}

\author{Thomas Wellens}
\affiliation{
 Fraunhofer Institute for Applied Solid State Physics, Tullastr. 72, 79108 Freiburg, Germany
}

\date{\today}

\begin{abstract}
A common approach to deal with gate errors in modern quantum-computing hardware is zero-noise extrapolation. By artificially amplifying errors and extrapolating the expectation values obtained with different error strengths towards the zero-error (zero-noise) limit, the technique aims at rectifying errors in noisy quantum computing systems. For accurate extrapolation, it is essential to know the exact factors of the noise amplification. In this article, we propose a simple method for estimating the strength of errors occurring in a quantum circuit and demonstrate improved extrapolation results. The method determines the error strength for a circuit by appending to it the inverted circuit and measuring the probability of the initial state. The estimation of error strengths is easy to implement for arbitrary circuits and does not require previous characterization of noise properties. We compare this method with the conventional zero-noise extrapolation method and show that our method leads to a more accurate calculation of expectation values on current quantum hardware, showcasing its suitability for near-term quantum computing applications.

\end{abstract}

\maketitle

\section{\label{sec:introduction}Introduction}
Current quantum computing devices, so-called Noisy Intermediate Scale Quantum (NISQ) hardware \cite{Preskill_2018}, suffer from errors \cite{Preskill_2018, Temme2017, Li2017, Bonet2018}. The accuracy of the output is reduced by these errors. Quantum error mitigation techniques aim to minimize these inaccuracies. The general principle of these techniques is to deduce an almost noise-free estimate of the expectation value of interest from results obtained on noisy hardware. For example, the technique of zero-noise extrapolation (ZNE) \cite{Temme2017, Kandala2019, Kim_2023, Li2017, Krebs2022, Bharti2022} consists of performing a series of measurements with systematically amplified errors. The expectation values obtained with different error strengths are then extrapolated to zero error (zero-noise). Hence, a more accurate estimation of the expectation value can be obtained at the cost of an increased sampling overhead due to a larger number of circuits and classical post-processing.

Several zero-noise extrapolation schemes have been proposed, each with their own advantages and challenges. In continuous ZNE \cite{Temme2017,Kandala2019}, the microwave pulses generating the elementary single- and two-qubit gates are scaled to increase the noise throughout the circuit. This extrapolation yields satisfactory results, however, it requires a time-consuming calibration of the scaled pulses. The implementation of the discrete ZNE method \cite{He2020, Dumitrescu2018, Giurgica2020, McCaskey2019} is more straightforward. Here, sequences of gates are inserted that, in the error-free case, are equivalent to the identity. Depending on the details of the implementation, different noise scaling factors can be targeted; for example, replacing every CNOT gate with three CNOT gates corresponds to a noise-scaling factor of three. While this simple scaling behavior holds, for example, for depolarizing errors to the limit of small error strength, this is usually not the case for realistic noise models. Depending on the specific type of errors occurring on the given hardware, not every error is amplified exactly $\lambda$ times if the corresponding gate is executed $\lambda$ times \cite{hour2024, majumdar2023,Schultz2022}. 
This may lead to a certain bias concerning the estimation of the error strength, which deteriorates the results of the discrete ZNE method \cite{hour2024, majumdar2023}. Another powerful approach is ZNE with Probabilistic Error Amplification (PEA)~\cite{Kim2023_util}.
With PEA, a noise model is first learned for different circuit layers, and errors are then amplified by inserting random gates according to the noise model. This approach has a high sampling cost due to the larger number of experiments required to learn the noise model.
Probabilistic Error Cancellation (PEC) \cite{Temme2017, Berg2023} is another method with an even higher sampling cost \cite{Mari2021} that represents the ideal circuit as a quasi-probability distribution of previously characterized noisy ones.
Finally, if the noise can be described by a depolarizing noise model, the expectation value can be corrected after measuring the strength of the depolarizing noise \cite{Urbanek_2021, Rahman_2022} using a method based on inverted circuits similar to the one which we will introduce below in a more general context. This correction is easy to implement, but it is only applicable in the case of errors well approximated by depolarizing noise.

In this article, by building on the relatively simple method of ZNE with identity insertions \cite{He2020}, we introduce the method of Inverted-Circuit-ZNE (IC-ZNE) which allows for the measurement of the entire error of a circuit to obtain an exact noise scaling parameter for better error mitigation. This involves adding an extra circuit with its inverse to determine the actual error strength $\epsilon$. This method overcomes the above problem that not every error type is amplified $\lambda$ times when increasing the number of noisy gates by $\lambda$ \cite{Kim_2023}. We further explicitly show that the total error is actually different from the assumed error amplification. Note that our IC-ZNE approach does not require any knowledge of the underlying quantum computer noise model.

Methods that also involve an inverted circuit are (dual-state) purification \cite{Cai_2021_puri, Huo_2022} and verified phase estimation \cite{OBrien_2021}. The main idea of dual-state purification is to consider $\rho^2/{\rm} Tr(\rho^2)$ as a purified version of the noisy quantum state $\rho$. When combined with ZNE (as in Refs.~\cite{yang_2024, Yoshioka_2022}), this method can be regarded as complementary to our IC-ZNE approach: concerning the expectation values obtained for different circuits with scaled noise, dual-state purification yields more precise (purified) expectation values, whereas our method focuses on the accurate determination of the corresponding noise strengths.\\
A more general framework (containing the above-mentioned dual-state purification as sub-version) is provided by the generalized quantum subspace expansion \cite{yang_2024, Yoshioka_2022}. Another sub-version (fault subspaces) is related to ZNE, since, similar as in ZNE, it considers superpositions of states created by circuits with different noise strengths. In contrast to ZNE, the coefficients of this superposition are obtained from a variational principle and not by extrapolation to zero noise. This method can be applied to variational problems (e.g., finding the ground state energy of a given Hamiltonian), where it appears to be well suited due to its robustness against inaccurate knowledge of noise levels \cite{Yoshioka_2022}. In contrast, our IC-ZNE approach can be used to mitigate expectation values for arbitrary observables and quantum circuits, not restricted to variational problems.\\
Finally, the verified phase estimation approach mentioned above uses inverted circuits to detect and discard errors \cite{OBrien_2021}. However, it can be applied only for the purpose of phase estimation (or more generally, if the prepared quantum state is an eigenstate of the measured observable).

The paper is organized as follows. After introducing known concepts and methods used in this paper like standard ZNE (sZNE) and  randomized compiling in Sec.~\ref{sec: Concepts and Methods}, we introduce the method of Inverted-Circuit ZNE (IC-ZNE) in Sec.~\ref{sec:IC-ZNE}. Starting from a precise definition of the error strength $\epsilon$, we derive a relation between $\epsilon$ and the probability of measuring all qubits in their initial state after the inverse circuit has been added.
Finally, we demonstrate and explain the improved performance of our method compared to standard ZNE on simulators with depolarizing noise and on IBM's quantum computing hardware in Sec. \ref{sec:results}.   
A conclusion is drawn in Sec. \ref{sec:conclusion}.

\section{\label{sec: Concepts and Methods}Concepts and Methods}
\subsection{General setting and notation}
\label{sec:general}

We consider the following scenario: The quantity of interest is given by the expectation value of observable $A$ with respect to a state prepared by a quantum circuit $U$
\begin{equation}
\langle A\rangle_{\rm ideal}=\langle \psi|A |\psi\rangle,
\label{eq:Aideal}
\end{equation}
where
\begin{equation}
|\psi\rangle = U|0\rangle \label{eq:psi}
\end{equation}
and $|0\rangle$ refers to the initial state.
Here, the symbol $U$ refers to the ideal unitary quantum operation. In actual quantum devices, however, a noisy quantum channel ${\mathcal E}_U$ is implemented instead of the targeted unitary operation $U$ with corresponding noisy expectation value
\begin{equation}
\langle A\rangle = {\rm tr}(A\rho)\label{eq:A},
\end{equation}
where
\begin{equation}
\rho = {\mathcal E}_U (|0\rangle\langle 0|). \label{eq:rho}
\end{equation}
The main idea of ZNE now consists of introducing a parameter $\epsilon$ which is expected to describe the amount of noise present in ${\mathcal E}_U$. Assuming that this parameter, or at least its scaling $\lambda = \epsilon/\epsilon_0$ with respect to the (in general unknown) noise strength $\epsilon_0$ of the original circuit, can be controlled by adding additional noise to the system, the data points $\langle A\rangle_{\lambda_i}$ measured at different noise scaling factors $\lambda_i$ can be extrapolated to yield
$\langle A\rangle_{\rm ideal}=\langle A\rangle_{\lambda\to 0}$. Different versions of ZNE differ in the way in which the noise parameter $\epsilon$ (or, equivalently, the noise scaling factor $\lambda$) is defined, the number of data points to be measured at different values of the noise strengths and the form of the function fitted to these points. 

\subsection{\label{sec:std-zne}Standard Zero-Noise Extrapolation}
A noise-model agnostic approach for scaling the noise strength is to replace unitary operations \textit{U} by $U(U^\dagger U)^n$ \cite{He2020, Giurgica2020}. The additional operation $U^\dagger U$ does not logically alter the circuit, since 
$U^\dagger U=I$
corresponds to the identity in the absence of noise. As the time needed to execute the sequence $U(U^\dagger U)^n$, however, is $2n+1$ times longer than for $U$ alone, one may expect that errors occurring in ${\mathcal E}_U$ are amplified by a factor 
\begin{equation}
\lambda=2n+1. 
\end{equation}
In particular, this is true if the noisy implementations ${\mathcal E}_U$ and ${\mathcal E}_{U^\dagger}$ of both, $U$ and $U^\dagger$, amount to a depolarizing error $\Lambda_p$ with error probability $p$, i.e.
\begin{eqnarray}
    {\mathcal E}_U & = & \Lambda_p {\mathcal U}
    \label{eq:U_depol}\\
    {\mathcal E}_{U^\dagger} & = & \Lambda_p {\mathcal U}^\dagger.
\end{eqnarray}
Here, ${\mathcal U}(\rho)=U\rho U^\dagger$ and ${\mathcal U}^\dagger(\rho)=U^\dagger\rho U$ denote the unitary quantum channels defined by $U$ and $U^\dagger$, respectively, and
\begin{equation}
    \Lambda_p(\rho) = (1-p)\rho + p \rho_*
    \label{eq:Lambda_depol}
\end{equation}
with maximally mixed state $\rho_*=I/2^q$ \cite{Nielsen2010} (where $I$ denotes the identity and $q$ the number of qubits).

In this case, the noisy implementation of the sequence $U(U^\dagger U)^n$ indeed yields again a depolarizing error with scaled error probability $\lambda p$ (in first order of $p$) \cite{He2020, Giurgica2020}. This can be traced back to the fact that depolarizing errors commute with unitary operations, i.e.,
$\Lambda_p {\mathcal U}={\mathcal U}\Lambda_p$ and $\Lambda_p {\mathcal U}^\dagger={\mathcal U}^\dagger\Lambda_p$, and hence
\begin{eqnarray}
    {\mathcal E}_U \left({\mathcal E}_{U^\dagger} {\mathcal E}_U\right)^n & = &  \Lambda_p {\mathcal U} \left(\Lambda_p {\mathcal U}^\dagger\Lambda_p {\mathcal U}\right)^n\nonumber\\
    & = & \Lambda_p^{2n+1} {\mathcal U} 
    \left({\mathcal U}^\dagger{\mathcal U}\right)^n\nonumber \\
    & = & {\Lambda_{1-(1-p)^{2n+1}}~{\mathcal U}}.
\end{eqnarray}
Indeed, we obtain the same result as in Eq.~(\ref{eq:U_depol}) with re-scaled probability of the depolarizing error, 
i.e., $1-p\to (1-p)^\lambda$, where $\lambda=2n+1$,
which simplifies to $p\to \lambda p$ for $\lambda p\ll 1$.
The expectation value of $A$ results as:
\begin{equation}
    \langle A\rangle =(1-p)^\lambda \langle A\rangle_{\rm ideal}+\left[1-(1-p)^\lambda\right] \langle A\rangle_*,
    \label{eq:exp_fit}
\end{equation}
where $\langle A\rangle_*={\rm tr}(A\rho_*)$, or
\begin{equation}
    \langle A\rangle \simeq (1-\lambda p)\langle A\rangle_{\rm ideal}+\lambda p\langle A\rangle_*
    \label{eq:linear_fit}
\end{equation}
for $\lambda p\ll 1$.

For other error models, however, a similarly simple scaling behavior generally does not hold. In the case of coherent errors, it may even happen that the errors of ${\mathcal E}_U$ and ${\mathcal E}_{U^\dagger}$ exactly compensate for each other, such that ${\mathcal E}_{U^\dagger}{\mathcal E}_{U}=I$ and the error is not amplified at all.

Due to its simplicity and moderate sampling overhead, the method of ZNE based on identity insertions is frequently used in order to mitigate gate errors. In this article, we will use it as a benchmark for our more refined method introduced in Sec.~\ref{sec:IC-ZNE}. More specifically, we employ the following version, subsequently referred to as \lq standard ZNE\rq\ (sZNE): first, we concentrate on two-qubit gate (i.e., CNOT) errors, which are the dominant source of errors in the superconducting device provided by IBM (i.e., one or two orders of magnitude larger than the errors of single-qubit gates) \cite{Chow2011, Sheldon_2016, Kandala_2021, Magesan_2011,Magesan_2012}. We use noise scaling factors $\lambda=1$ (the original circuit), $\lambda=3$ (every CNOT gate replaced by 3 CNOT gates) and $\lambda=5$ (5 CNOT gates) and fit the measured data points by an
exponential function $f(\lambda)=a_1e^{-a_2\lambda}+a_3$ motivated by
Eq.~(\ref{eq:exp_fit}), which is generally expected to yield more accurate results than a linear fit  based on Eq.~(\ref{eq:linear_fit}) \cite{Endo2018}. 
When performing the exponential fit, we restrict the parameters $a_1$ and $a_3$ to the interval $[A_{\rm min},A_{\rm max}]$ given by the minimum and maximum eigenvalue of $A$.
Without this restriction, we have observed that the exponential fit may be highly unstable in some cases, especially if noise strengths are relatively large.
Furthermore,
in order to evaluate the expectation value $\langle A\rangle$, we employ readout error mitigation based on the method M3 \cite{Nation2021}.

\subsection{\label{sec:additional techniques}Randomized compiling}

\begin{figure}
    \includegraphics[scale=.09]{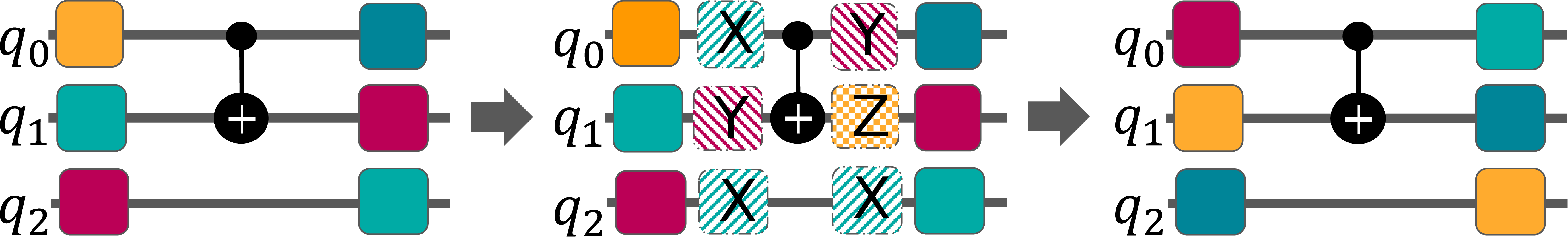}
  \caption{\label{fig:twirlb} On the left: Example of a bare circuit with single qubit gates and one CNOT gate. In the middle: Twirling is performed by adding random Pauli gates (rectangles with hatching) to the circuit without changing the logic of the CNOT gates in the circuit. Twirling is only applied before and after CNOT gates, as well as to adjacent qubits, which could be affected by the CNOT gate due to crosstalk. Finally, the number of single qubit gates is reduced by contraction of subsequent single-qubit gates (circuit on the right).}
\end{figure}

Besides readout error mitigation, additional techniques can be used to enhance the performance of ZNE \cite{Hashim2021, Kim_2023, majumdar2023, Kurita2023, Cai2021, cai2023practical}. In this paper, we will investigate the effect of randomized compiling or so-called Pauli twirling \cite{Wall2016} on ZNE. 
The technique of Pauli twirling is based on the fact that the CNOT gate is a Clifford gate \cite{Gottesman1998}, and hence every Pauli operator is transformed into a Pauli operator again, after sandwiching it between two CNOT operations:
\begin{equation}
{\rm CNOT}~{\mathcal P}_i~{\rm CNOT} = {\mathcal P}_{\pi(i)},
\label{eq:Pauli_CNOT_commutation}
\end{equation}
where ${\mathcal P}_i$, $i=1,2,\dots,4^q$, refer to multi-qubit Pauli operators (i.e., tensor products of the single-qubit Pauli operators $I,X,Y$ or $Z$ acting on each qubit) and $\pi$ denotes the corresponding permutation. Since ${\rm CNOT}^\dagger={\rm CNOT}$ and ${\mathcal P}_i^2=I$, this can be rewritten as:
\begin{equation}
{\mathcal P}_i~{\rm CNOT}~{\mathcal P}_{\pi(i)} = {\rm CNOT}.
\end{equation}
Sandwiching each CNOT gate between two Pauli operators ${\mathcal P}_i$ and ${\mathcal P}_{\pi(i)}$ with randomly chosen $i$ hence preserves the CNOT logic in the absence of noise.
In the presence of noise, however, the error channel $\Lambda$ of the noisy CNOT gate ${\mathcal E}_{\rm CNOT}=\Lambda~{\rm CNOT}$ is transformed, on average, into a Pauli channel $\Lambda_\mathcal{P}$ of the following form:
\begin{equation}
\Lambda_{\mathcal P}(\rho)=\sum_{i=1}^{4^q} p_i {\mathcal P}_i \rho {\mathcal P}_i,\ \ p_i>0,\ \ \sum_{i=1}^{4^q} p_i=1.
\end{equation}
In other words, the Pauli operator ${\mathcal P}_i$ is applied with probability $p_i$, and thus produces an error if ${\mathcal P}_i\neq I$.
Twirling is also applied to adjacent qubits, which could suffer from crosstalk \cite{Zhao2022, Rudinger2021}, as described in \cite{ketterer2023}. Each of the twirling gates is combined with other single-qubit operations to avoid increasing circuit length and therefore time (see Fig. \ref{fig:twirlb}). Note that if the noisy CNOT gate originally suffers from a coherent error (i.e., if its error channel $\Lambda$ is unitary), the latter is transformed into an incoherent error by Pauli twirling. As already hinted at in Sec.~\ref{sec:std-zne}, the scaling behavior of the noise due to repeated CNOT gates is expected to be more accurate in this case. 

\section{\label{sec:IC-ZNE}Inverted-Circuit Zero-Noise Extrapolation}
As explained in Sec.~\ref{sec:std-zne}, the standard ZNE relies on a certain assumption concerning the scaling of the error strength, namely that the error of a gate is amplified by a factor $\lambda$ if the gate is repeated $\lambda-1$ times. In general, however, this is only true for small errors commuting with the ideal gate operation (such as depolarizing errors).
We therefore propose not to rely on this assumption and, instead, to measure the error strength $\epsilon$ of a quantum circuit $U$ used to prepare state $|\psi\rangle$ directly.    

\subsection{\label{sec:level3}Definition of error strength $\epsilon$}

Let us consider the fidelity $F$ of the noisy state $\rho$, see Eq.~(\ref{eq:rho}), with respect to the ideally prepared state $|\psi\rangle$, see Eq.~(\ref{eq:psi}):
\begin{equation}
F = \langle \psi|\rho|\psi\rangle
\label{eq:F},
\end{equation}
and define
\begin{equation}
\epsilon = 1-F.
\label{eq:p}
\end{equation}
In order to derive the implications of this definition for ZNE, we
decompose the density matrix as follows:
\begin{equation}
\rho = (1-\epsilon) |\psi\rangle\langle\psi| + \epsilon \sigma.
\label{eq:rho_decomp}
\end{equation}
The operator $\sigma$ represents the noisy part. 
In case of depolarizing errors, we obtain $\sigma=(I-|\psi\rangle\langle\psi|)/(2^q-1)$ (independent of $\epsilon$) and the following relation between the noise strength $\epsilon$ and the error probability $p$ of the depolarizing channel: $\epsilon=p\left(1-\frac{1}{2^q}\right)$.
In general, however, $\sigma$ and $\epsilon$ are, a priori, unknown. 
Note that $\sigma$ is not even guaranteed to be positive, but this poses no problem in the following (where we will not assume that $\sigma$ is positive). From Eqs.~(\ref{eq:F}-\ref{eq:rho_decomp}), it follows that:
\begin{equation}
\langle\psi|\sigma|\psi\rangle=0.
\end{equation}
The expectation value of an observable $A$ can then be written as follows:
\begin{equation}
\langle A\rangle = (1-\epsilon) \underbrace{\langle\psi|A|\psi\rangle}_{=\langle A\rangle_{\rm ideal}}+\epsilon~{\rm tr}(A\sigma).
\label{eq:A_epsilon}
\end{equation}

It converges to $\langle A\rangle_{\rm ideal}$, see Eq.~(\ref{eq:Aideal}), for $\epsilon\to 0$.
Moreover, note that, due to our definition of  $\epsilon$ based on the fidelity, cf. Eq.~\eqref{eq:p}, the error strength $\epsilon$ is restricted to the interval $[0,1]$ -- in contrast to the noise scaling factor $\lambda$ based on the number of gate repetitions, which may, in principle, be chosen arbitrarily large. For this reason,  
a linear fit of $\langle A\rangle$ as a function of $\epsilon$ should be sufficiently accurate for IC-ZNE,
as is evident from Eq.~(\ref{eq:A_epsilon}),
at least if the dependence of the noisy part $\sigma$ on $\epsilon$ is negligible. Regarding the latter point, the technique of randomized compiling may be useful (refer to Sec. ~\ref{sec:additional techniques}) as it transforms coherent errors into incoherent ones.

\subsection{\label{sec:level4}Measuring the error strength}

\begin{figure}
    \includegraphics[width=0.45\textwidth]{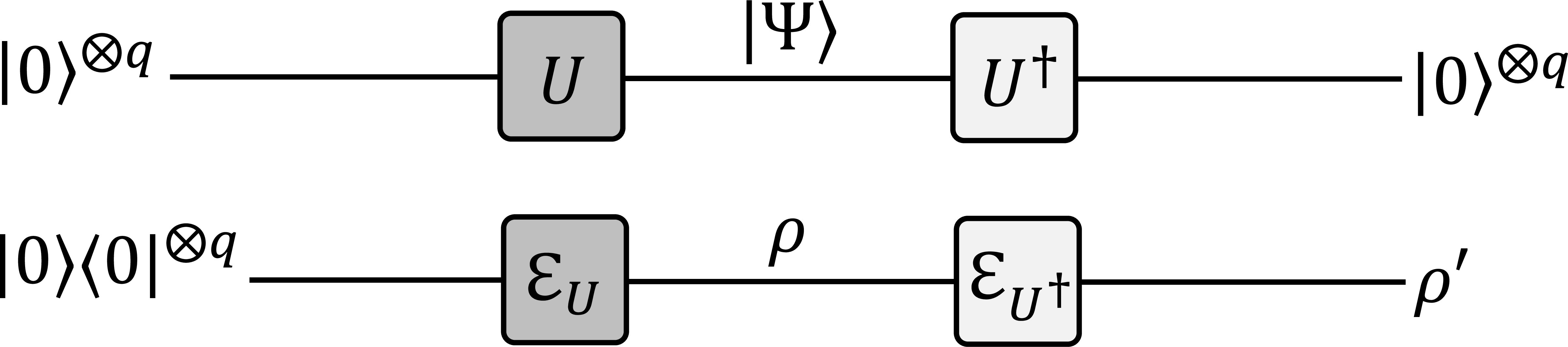}
  \caption{\label{fig:iczne_circ}Schematic of the IC-ZNE method. Without noise, adding the inverse circuit $U^\dagger$ after $U$ restores the initial state $|0\rangle$. 
  The corresponding noisy channels ${\mathcal E}_U$ and ${\mathcal E}_{U^\dagger}$, however, generate a density matrix $\rho'$, which, as shown in the main text, contains information about the overall error strength in the circuit.}
\end{figure}

After having defined the error strength $\epsilon$, we now show how it can be measured for a given circuit $U$. The basic idea is the following: We apply first $U$, then $U^\dagger$ and measure the probability $P_0$ to find all qubits in state $|0\rangle$, which is the initial state.
Without noise, we obviously obtain $P_0=1$.
The corresponding noisy operations ${\mathcal E}_U$ and ${\mathcal E}_{U^\dagger}$ (see lower part in Fig.~\ref{fig:iczne_circ}), however, yield the following result:
\begin{equation}
    P_0 = \langle 0|\rho'|0\rangle,
    \label{eq:p00}
\end{equation}
where
\begin{equation}
\rho'=({\mathcal E}_{U^\dagger}{\mathcal E}_U)\left(|0\rangle\langle 0|\right).
\label{eq:rhostr}
\end{equation}
Substituting Eq.~(\ref{eq:rhostr}) in Eq.~(\ref{eq:p00}) yields:
\begin{eqnarray}
P_0 & = & \Bigl< 0\Bigl|({\mathcal E}_{U^\dagger}{\mathcal E}_U)\left(|0\rangle\langle 0|\right)\Bigr|0\Bigr>\nonumber\\
& = & {\rm tr}\bigl\{|0\rangle\langle 0|~({\mathcal E}_{U^\dagger}{\mathcal E}_U)\left(|0\rangle\langle 0|\right)\}\nonumber\\
& = & {\rm tr}\bigl\{{\mathcal E}^\dagger_{U^\dagger}\left(|0\rangle\langle 0|\right)~ {\mathcal E}_U\left(|0\rangle\langle 0|\right) \bigr\}\nonumber\\
& = & {\rm tr}\left\{\tilde{\rho}\rho\right\}\label{eq:P0},
\end{eqnarray}
where
\begin{equation}
\tilde{\rho} = {\mathcal E}_{U^\dagger}^\dagger\left(|0\rangle\langle 0|\right).
\end{equation}
Here, ${\mathcal E}_{U^\dagger}^\dagger$ denotes the adjoint quantum channel, generally defined as ${\rm tr}\left[A~{\mathcal E}(B)\right]={\rm tr}\left[{\mathcal E}^\dagger(A) B\right]$.
The state $\tilde{\rho}$ is called "dual state" in Refs.~\cite{Cai_2021_puri, Huo_2022}, and the same circuit that we propose in order to measure the error strength has been used there to determine the purity ${\rm Tr}(\rho \tilde{\rho})$, see Eq.~\eqref{eq:P0}, needed to measure the purified expectation value $\langle A\rangle={\rm Tr}\{A\rho\tilde{\rho}\}/{\rm Tr}\{\rho \tilde{\rho}\}$.

In case of a depolarizing error channel (which is self-adjoint and commutes with unitary channels), we obtain $\tilde{\rho}=\rho$, but, 
in general, the operator $\tilde{\rho}$ differs from $\rho$. However, we assume that it exhibits the same fidelity as $\rho$, see Appendix~\ref{sec:level5} for justifying this assumption. Then, we can decompose $\tilde{\rho}$ in the same way as $\rho$:
\begin{equation}
\tilde{\rho} = (1-\epsilon) |\psi\rangle\langle\psi| + \epsilon \tilde{\sigma},
\label{eq:rhotilde_decomp}
\end{equation}
where $\langle\psi|\tilde{\sigma}|\psi\rangle=0$. Using the decomposition of $\rho$ and $\tilde{\rho}$ in Eqs.~(\ref{eq:rho_decomp}, \ref{eq:rhotilde_decomp}) together with $\langle\psi|\sigma|\psi\rangle=\langle\psi|\tilde{\sigma}|\psi\rangle=0$, we arrive at:
\begin{equation}
P_0 = (1-\epsilon)^2+\epsilon^2 \underbrace{{\rm tr}(\sigma\tilde{\sigma})}_{=: a}\label{eq:P02},
\end{equation}
where we define the quantity $a$ depending on the operators $\sigma$ and $\tilde{\sigma}$.

\begin{figure}[]
    \includegraphics[width=0.5\textwidth]{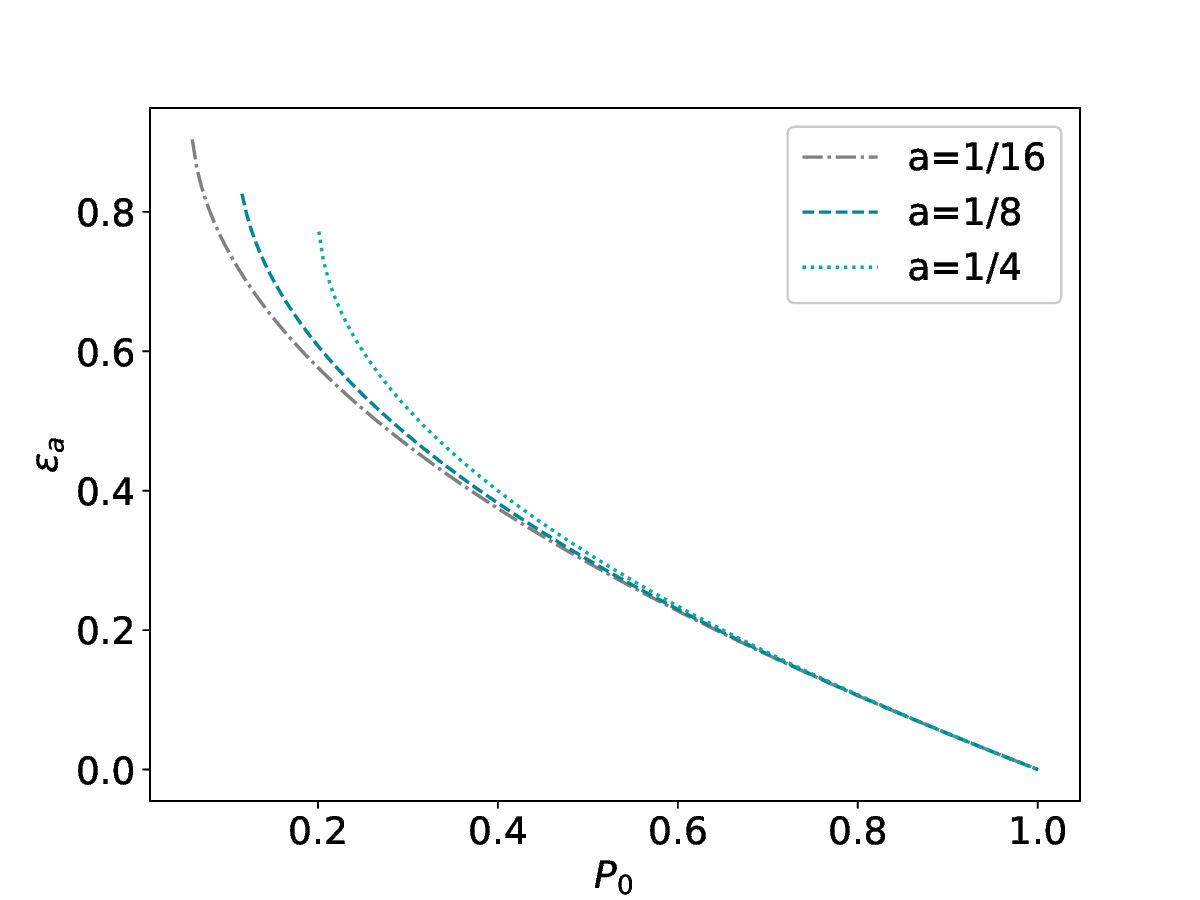}
  \caption{\label{fig:poverp0} Noise strength $\epsilon_a$, see Eq.~(\ref{eq:epsilon_a}), as a function of the probability $P_0$ to measure every qubit in state $|0\rangle$ for different values of $a$, see Eq.~(\ref{eq:P02}). In a low-noise regime, the choice of $a$ has no major effect on the value of the error strength.}
\end{figure}
 
The noise strength $\epsilon$ can now be deduced from the measured quantity $P_0$ by solving Eq.~(\ref{eq:P02}) for $\epsilon$. If $P_0>a$, there is a unique solution with $0\leq \epsilon\leq 1$:
\begin{equation}
\epsilon_a = \frac{1-\sqrt{P_0-a(1-P_0)}}{1+a},
\label{eq:epsilon_a}
\end{equation}
where $\epsilon_a$ denotes the dependence of $\epsilon$ on $a$.
If $\frac{a}{1+a}<P_0\leq a$, a second solution is found, whereas for $P_0<\frac{a}{1+a}$, no solution of Eq.~(\ref{eq:P02}) exists. These issues will be discussed in the following section.

First, we note that the quantity $a$ is unknown, but expected to be small for incoherent errors, where the purity of $\sigma$ and $\tilde{\sigma}$ is low. 
Indeed, if the gate errors are described by Pauli channels (which, for CNOT gates, can be enforced by randomized compiling, see Sec.~\ref{sec:additional techniques}), then $\sigma$ and $\tilde{\sigma}$ arise from a mixture of a large number of states originating from all possible errors occurring at each gate. Thus, $a={\rm tr}(\sigma \tilde{\sigma})\leq \frac{1}{2}[{\rm tr}(\sigma^2)+{\rm tr}(\tilde{\sigma}^2)]\ll 1$. 
In case of a depolarizing error, e.g., we obtain $a=\frac{1}{2^q-1}$ where $q$ denotes the number of qubits.

If we now plot $\epsilon_a$ as a function of $P_0$ (see Fig. \ref{fig:poverp0}), we see that, especially in the regime of low noise ($P_0$ close to 1, $\epsilon_a$ close to 0), the exact choice of $a\ll 1$ is not important, since 
\begin{equation}
\epsilon_a\to \frac{1-P_0}{2} \text{ for } P_0\to 1
\label{eq:factor_two}
\end{equation}
independent of $a$. Quite intuitively, the factor two in the denominator can be explained by the fact that the error of ${\mathcal E}_U$ is multiplied by a factor of two after adding ${\mathcal E}_{U\dagger}$. Note that Eq.~(\ref{eq:factor_two}) is not true for coherent errors, where the assumption $a\ll 1$ is not fulfilled. (In this case, we have checked that $a\propto \frac{1}{\epsilon}$ for $\epsilon\to 0$.)

Coming back to the incoherent case, we take into account deviations from the asymptotic behavior, Eq.~(\ref{eq:factor_two}), by assuming a reasonable value of $a$, e.g. $a=\frac{1}{2^q}$ (slightly larger than the result obtained for depolarizing error, see above). If $P_0<\frac{1}{2^q}$, we set $a=P_0$ to avoid the cases of no solution or two solutions mentioned above. In total, we finally obtain the following estimate of the noise strength $\epsilon$ from the measured probability $P_0$ of detecting all $q$ qubits in state $|0\rangle$:
\begin{equation}
\epsilon = \left\{\begin{array}{cc}
\frac{1-\sqrt{P_0-\frac{1-P_0}{2^q}}}{1+\frac{1}{2^q}}
& P_0>\frac{1}{2^q}\\
& \\
\frac{1-P_0}{1+P_0} & P_0\leq \frac{1}{2^q}.\end{array}
\right.
\end{equation}

\subsection{\label{sec:iczne_implementation}Structure of IC-ZNE application}

The method IC-ZNE consists of the following steps: first, we take the same circuits which we also use in standard ZNE, i.e., with noise scaling factors $\lambda=1,3$ and $5$ obtained from repeated CNOT gates, and measure the expectation value $\langle A\rangle$ for each of them. Second, the noise strength $\epsilon$ of each circuit is measured by adding its inverse as explained above. Finally, we plot $\langle A\rangle$ as a function of $\epsilon$ and extrapolate to $\epsilon\to 0$ using a linear fit. More details and examples will be shown in Sec.~\ref{sec:fitting_procedure}. It is noteworthy that any noise scaling factor can be chosen for IC-ZNE in principle.

\section{\label{sec:results}Benchmarking Inverted-Circuit Zero-Noise Extrapolation}
\subsection{\label{sec:circuits}Sample circuits}

In this section, we will test standard ZNE and IC-ZNE with two different circuits on simulators and on quantum hardware. 
The first circuit is an example of  Grover's quantum search algorithm \cite{grover1996} with three qubits and 10 CNOT gates (after specifically tailoring it to conform to IBM’s quantum device architecture, here \textit{ibmq\_ehningen}, through transpilation, see Appendix~\ref{sec:transpiled_circuits}). 
We consider the observable 
\begin{equation}
    A_{\rm Grover}=|101\rangle \langle 101|+|011\rangle \langle 011|
    \label{eq:ew_grover}
\end{equation}
corresponding to the probability of measuring one of the two solutions 101 or 011 of the search problem encoded by the oracle of our circuit. Without errors, we therefore obtain $\langle A_{\rm Grover}\rangle_{\rm ideal}=1$.

The second circuit is taken from the Harrow-Hassidim-Lloyd algorithm (HHL algorithm) \cite{Harr2009}. This algorithm can be used to solve linear systems of equations defined by a matrix $B\in\mathbb{C}^{N\times N}$ and a vector $\vec{b}\in\mathbb{C}^{N}$ to find $\vec{x}\in\mathbb{C}^{N}$, so that $B\Vec{x}=\Vec{b}$. We use an example with $N=2$ as described in \cite{HHL-IBM} and
\begin{equation}
    B=\begin{pmatrix}1 & -1/3\\-1/3 & 1 \end{pmatrix},\quad 
    \vec{b}=\begin{pmatrix}1 \\ 0\end{pmatrix}.
\end{equation}
The corresponding transpiled circuit for four qubits exhibits 18 CNOT gates and can be found in Appendix~\ref{sec:transpiled_circuits}. Only the last qubit is measured, and the corresponding expectation value of the observable 
\begin{equation}
        A_\text{HHL} =\mathbb{I} \otimes \mathbb{I} \otimes \mathbb{I} \otimes |1\rangle\langle 1|
        \label{eq:ew_hhl}
\end{equation}
i.e., the probability to measure the last qubit in state $|1\rangle$,
yields the norm of $\vec{x}$ 
as follows:
\begin{equation}
||\vec{x}||=\frac{3}{2}\sqrt{\langle A\text{HHL}  \rangle_{\rm ideal}}.
\end{equation}
In our case, $\langle A\text{HHL}  \rangle_{\rm ideal}=5/8$, in accordance with the correct solution $\vec{x}=(9/8,3/8)$ of the above linear system.

\subsection{Fitting procedure and noise scaling}
\label{sec:fitting_procedure}

In all cases (unless otherwise specified), we use the following parameters: the original circuits corresponding to $\lambda=1$ are the transpiled circuits shown in Appendix~\ref{sec:transpiled_circuits}. We generate two scaled circuits with scaling factors  $\lambda=3$ and 5 by multiplying CNOT gates, as explained in Sec.~\ref{sec:std-zne}. If randomized compiling is employed, 16 different twirled versions of each circuit are randomly generated and executed with $10~000/16=625$ shots, from which the expectation values $\langle A\rangle$ are determined separately for each of the 16 twirled versions. In the case of IC-ZNE, the noise strength $\epsilon$ of each version is additionally determined, again using  $625$ shots for every twirled circuit. The total number of shots thus amounts to 30~000 for sZNE (i.e., 10~000 shots per scaling factor $\lambda$) and 60~000 for IC-ZNE (due to the additional measurement of the error strength $\epsilon$). Thereby, we obtain in total $3\times 16=48$ data points $(\lambda,\langle A\rangle)$ (sZNE) or $(\epsilon,\langle A\rangle)$ (IC-ZNE), which are fitted by an exponential function or first-order polynomial to obtain the extrapolated noise-free value of $\langle A\rangle$ at $\lambda=0$ (sZNE) or $\epsilon=0$ (IC-ZNE), respectively.

\begin{figure*}%[ht]
    \includegraphics[width=\textwidth]{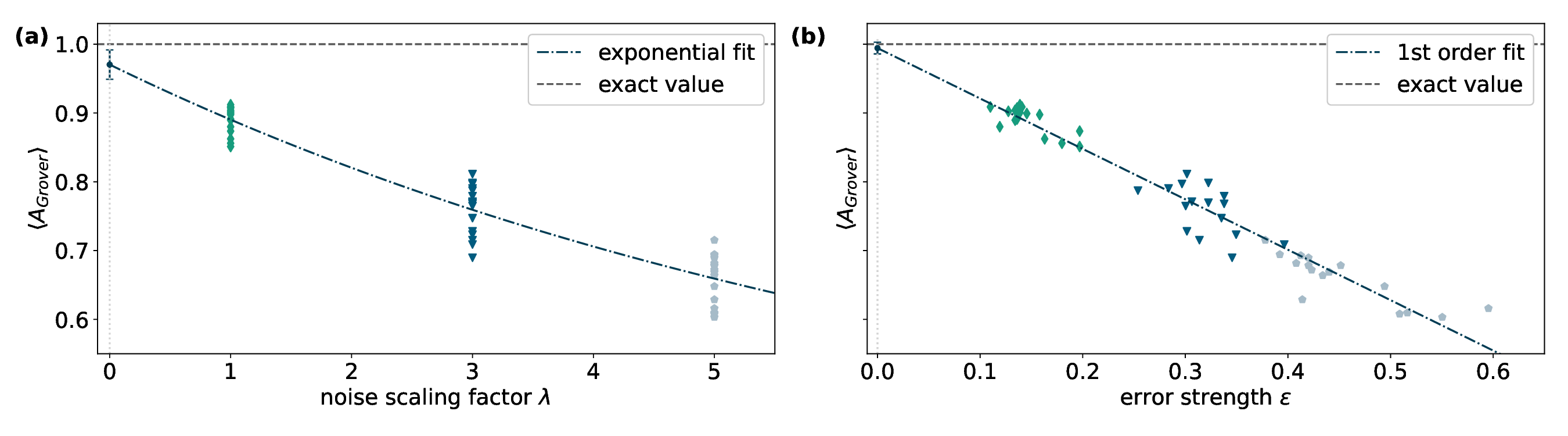}
    \caption{Standard ZNE (a) and Inverted-Circuit ZNE (b) for the Grover circuit performed on IBM's quantum device \textit{ibmq\_ehningen}. For each noise scaling factor $\lambda=1$ (green diamonds), $\lambda=3$ (blue triangles) and $\lambda=5$ (grey pentagons), we execute 16 different randomly twirled quantum circuits, each with 625 shots (amounting to 10~000 shots per scaling factor). The observed variation in the data points for the different noise scaling factors can be attributed to the use of different twirling gates, as well as to statistical fluctuations and variations introduced by the employed backend. The error bars (standard deviation) of the extrapolated values at zero noise ($\lambda=0$ or $\epsilon=0$, respectively) are determined from the covariance matrix of the fit parameters returned by the curve fitting routine.(a) In standard ZNE, the measured expectation values $\langle A_{\rm Grover}\rangle$ are plotted against the noise scaling factor $\lambda$. An exponential fit yields the extrapolated value $\langle A_{\rm Grover}\rangle=0.9703 \pm 0.02131$, which is significantly smaller than the ideal value $\langle A_{\rm Grover}\rangle_{\rm ideal}=1$ (horizontal dotted line). (b) In Inverted-Circuit ZNE, the error strength $\epsilon$ of each circuit is measured using the method of inverted circuits (with 625 additional shots per circuit). The same expectation values $\langle A_{\rm Grover}\rangle$ as in (a) are now plotted as a function of $\epsilon$ instead of $\lambda$, yielding a more accurate extrapolation result of $\langle A_{\rm Grover}\rangle=0.9943 \pm 0.0085$ with a first order linear fit.}
    \label{fig:grover_std_vs_iczne}
\end{figure*}

\begin{figure*}%[ht]
    \includegraphics[width=\textwidth]{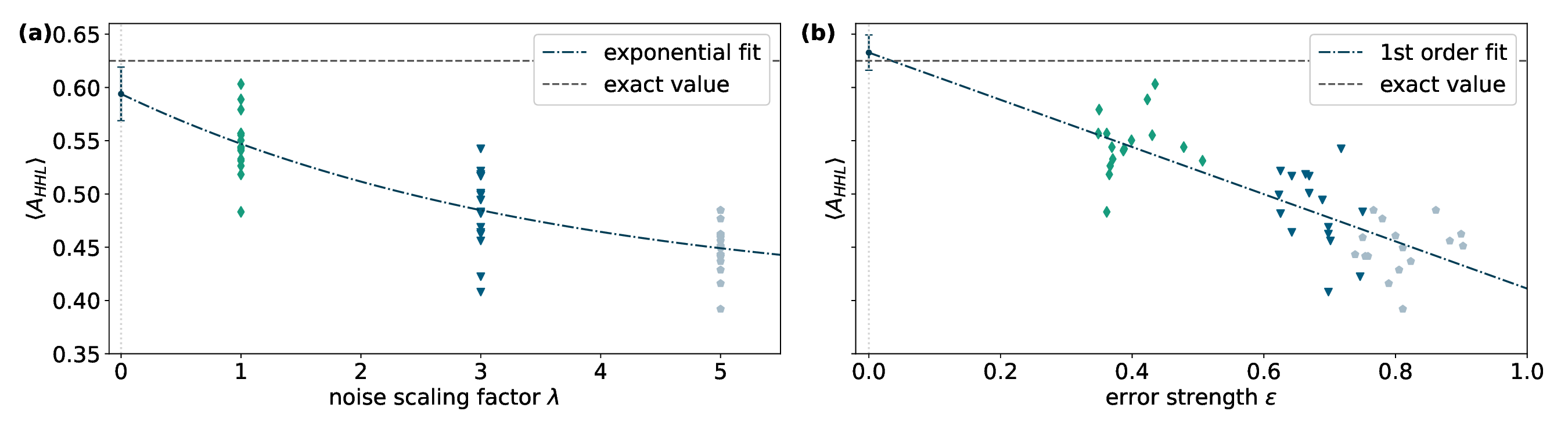}
    \caption{Same as Fig.~\ref{fig:grover_std_vs_iczne}, but for the HHL circuit instead of the Grover circuit. Again, the extrapolation using Inverted-Circuit ZNE (b) yields better agreement of the extrapolated value $\langle A_{\text{HHL} }\rangle$ $=0.5940 \pm  0.0252$ for sZNE and $0.6327 \pm 0.0166$ for IC-ZNE) with the exact value ($5/8=0.625$) than standard ZNE (a). Note that, due to a larger number of CNOT gates, the measured values of the error strengths $\epsilon$ are larger than in case of the Grover circuit, see Fig.~\ref{fig:grover_std_vs_iczne}b).}
    \label{fig:hhl_std_vs_iczne}
\end{figure*}

\begin{figure}
        \includegraphics[width=0.5\textwidth]{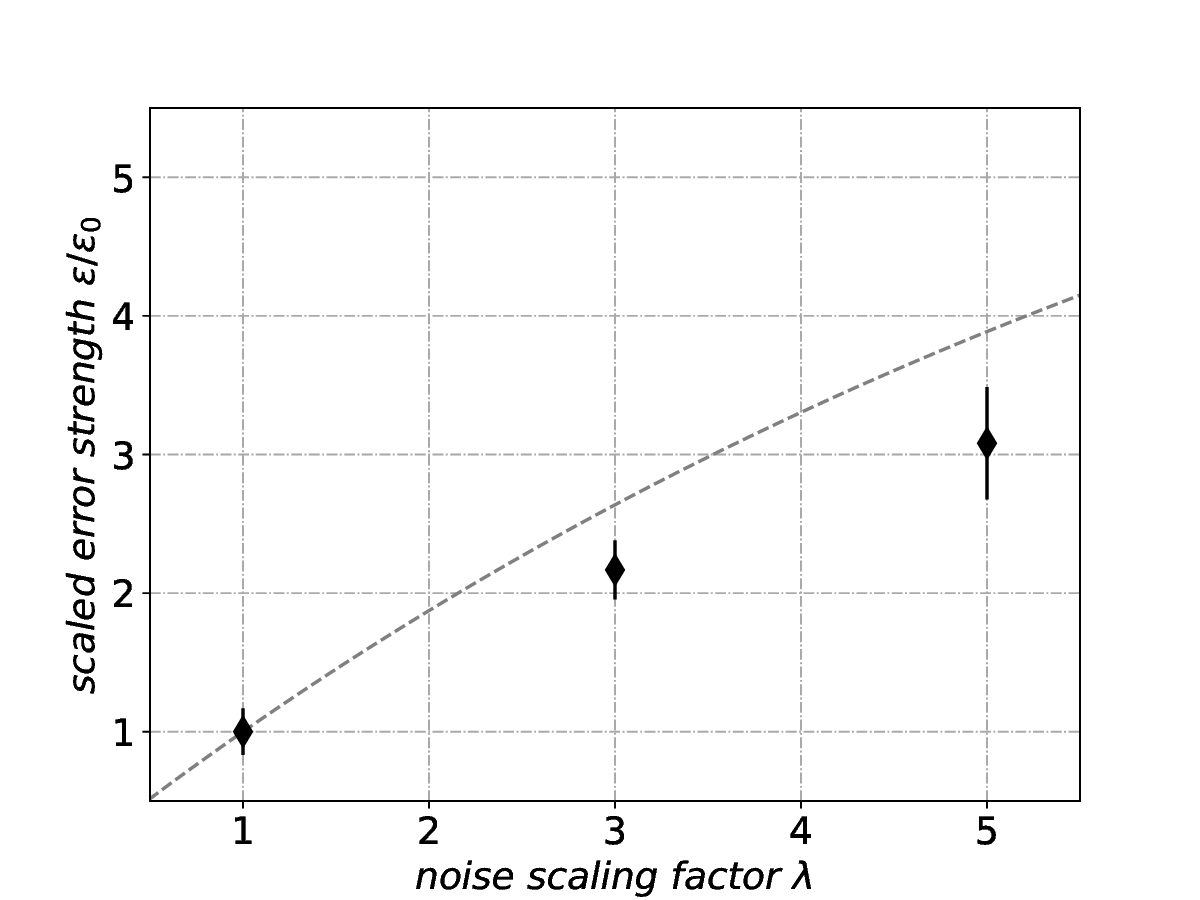}
    \caption{\label{fig:calcvstheor}Scaled error strength $\epsilon/\epsilon_0$ (mean values and standard deviation obtained from 16 randomly twirled circuits) as a function of the noise scaling factor $\lambda$ for the same data as in Fig.~\ref{fig:grover_std_vs_iczne} (Grover). The scaled error strength is normalized such that the mean value $\langle\epsilon\rangle/\epsilon_0=1$ for $\lambda=1$. A clear deviation from exponential scaling (dashed line), see Eq.~(\ref{eq:scaling}) with  $a_2=0.1350$ extracted from the exponential fit in Fig.~\ref{fig:grover_std_vs_iczne}(a),   
    assumed in standard ZNE is observed.}
\end{figure}

In Figs.~\ref{fig:grover_std_vs_iczne} and \ref{fig:hhl_std_vs_iczne}, the above procedure is illustrated for the case of the Grover and the HHL circuit, respectively, executed on IBM's device \textit{ibmq\_ehningen}. The corresponding transpiled circuits are displayed in Appendix~\ref{sec:transpiled_circuits}, whereas the average error rates of the individual CNOT gates, as obtained from the calibration data given by IBM, are shown in Appendix~\ref{sec:supplement}.
From Figs.~\ref{fig:grover_std_vs_iczne} and \ref{fig:hhl_std_vs_iczne}, we see that, in both cases, IC-ZNE yields a more accurate extrapolation than standard ZNE:
the extrapolated values at zero noise ($\lambda=0$ or $\epsilon=0$, respectively) are closer to the exact result, and their standard deviations (error bars) are smaller.
This can be explained by the fact that the error strengths of the scaled circuits deviate from the simple scaling behavior assumed in standard ZNE (see Sec.~\ref{sec:std-zne}).
To illustrate the deviation from the simple scaling behavior, we plot in Fig.~\ref{fig:calcvstheor}, for the same data as shown in Fig.~\ref{fig:grover_std_vs_iczne} (Grover circuit), the scaling of the error strengths 
$\epsilon/\epsilon_0$
determined by the method of inverted circuits as a function of the noise scaling factor $\lambda$,
where $\epsilon_0$ denotes the original error strength 
of the unscaled circuit ($\lambda=1$).
Remember that the factor $\lambda$ defines the scaled circuits by replacing each CNOT gate with $\lambda$ CNOT gates.
Standard ZNE with exponential fit assumes the following behavior of $\langle A\rangle$ as a function of $\lambda$:
\begin{equation}
    \langle A\rangle = a_1 e^{-a_2\lambda}+a_3
    \label{eq:exp_fit2}
\end{equation}
(see also Sec.~\ref{sec:std-zne}), where the
ideal value is reached at $\lambda=0$, i.e., $\langle A\rangle_{\rm ideal}=a_1+a_3$. Therefore, we can rewrite Eq.~(\ref{eq:exp_fit2}) as follows:
\begin{equation}
    \langle A\rangle = e^{-a_2\lambda}\langle A\rangle_{\rm ideal}+\left(1-e^{-a_2\lambda}\right)\langle A\rangle_{\rm noisy},
    \label{eq:exp_fit3}
\end{equation}
where $\langle A\rangle_{\rm noisy}=a_3$ refers to the expectation value reached in the strong-noise limit ($\lambda\to\infty$). This equation can be interpreted as follows: the quantum state at noise scaling factor $\lambda$ is a mixture of the ideal state $\rho_{\rm ideal}=|\psi\rangle\langle\psi|$ and a noisy state $\rho_{\rm noisy}$ (reached in the limit of strong noise):
\begin{equation}
    \rho = e^{-a_2\lambda}|\psi\rangle\langle\psi|+\left(1-e^{-a_2\lambda}\right)\rho_{\rm noisy}.
    \label{eq:rho_std_zne}
\end{equation}
In particular, this is true for depolarizing noise, where $\rho_{\rm noisy}=\rho_*$, see Sec.~\ref{sec:std-zne}.
Finally, Eq.~(\ref{eq:rho_std_zne}) yields the following relation between the noise scaling factor $\lambda$ and the error strength $\epsilon$, see Eqs.~(\ref{eq:F},\ref{eq:p}):
\begin{equation}
    \epsilon=\Bigl(1-\langle\psi|\rho_{\rm noisy}|\psi\rangle\Bigl)~\Bigl(1-e^{-a_2\lambda}\Bigr)
\end{equation}
or 
\begin{equation}
    \frac{\epsilon}{\epsilon_0} = \frac{1-e^{-a_2\lambda}}{1-e^{-a_2}},
    \label{eq:scaling}
\end{equation}
which we plotted in Fig.~\ref{fig:calcvstheor} (dashed line) as the scaling behaviour assumed in standard ZNE.

In contrast, the error strengths measured by the method of inverted circuits deviate significantly from this simple exponential scaling. 
This proves that the noise on the real quantum hardware cannot be adequately described by a depolarizing noise model (for which exponential scaling holds as shown above) and
demonstrates the benefits of IC-ZNE, which replaces the exponential scaling assumption with measured error strength values.

\subsection{Simulations with depolarizing noise}
\label{sec:simulation}

So far, we have presented the results of a single extrapolation run for both, the Grover and HHL circuits. To confirm the robustness of our conclusions, we will now present the results of a statistical analysis obtained from repeating the above procedure 50 times. We will consider, both: runs performed on real quantum hardware (see Sec.~\ref{sec:benchmark} below) as well as simulations with depolarizing noise of varying strength.

Regarding the latter, the CNOT and single-qubit gates are simulated using the depolarizing noise model. For simplicity, we assume that state preparation, and readout are error-free. This allows us to isolate the main source of error, which originates from CNOT gates in IBM's superconducting devices. We examine the effect of increasing noise strength by choosing values of 0.5\%, 1\%, 2\% and 5\% error rate per CNOT gate and 1/10 
of the respective value
for single qubit gates.

As discussed in Sec.~\ref{sec:fitting_procedure}, exponential scaling of the noise strength, as assumed in standard ZNE, can be proven to hold for fully depolarizing noise, where each noisy gate exhibits the same error channel given by Eq.~(\ref{eq:Lambda_depol}) with constant $p$.

\begin{figure*}%[htb]
    \includegraphics[width=\textwidth]{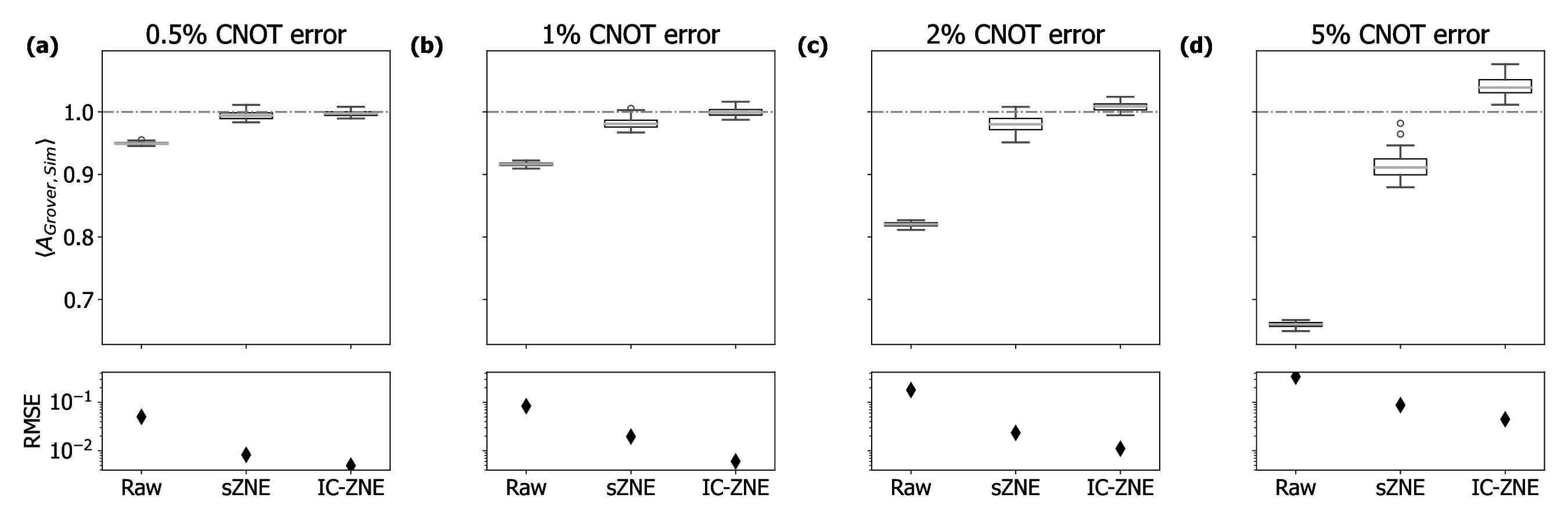}
    \caption{
    Comparison of inverted-circuit ZNE (IC-ZNE), standard ZNE (sZNE) and unmitigated raw data (Raw) for 50 simulation runs of the Grover circuit using a depolarizing error model with different error rates per CNOT gate: (a) 0.5\%, (b) 1\%, (c) 2\% and (d) 5\%.
   The error rate for single qubit gates is one-tenth of the CNOT error rate.
    For each method (Raw, sZNE and IC-ZNE), the obtained extrapolated values of $\langle A_{\rm Grover}\rangle$ are summarized in a box plot (see main text). The exact value $\langle A_{\rm Grover}\rangle_{\rm ideal}$ is shown by a dashed horizontal line. The root mean square error (RMSE), indicating the deviation from the exact value, is shown in the accompanying plot below on a logarithmic scale. In all cases, the inverted-circuit ZNE (IC-ZNE) exhibits the smallest root mean square error.
    }
    \label{fig:grover_simulation}
\end{figure*}

\begin{figure*}%[htb]
    \includegraphics[width=\textwidth]{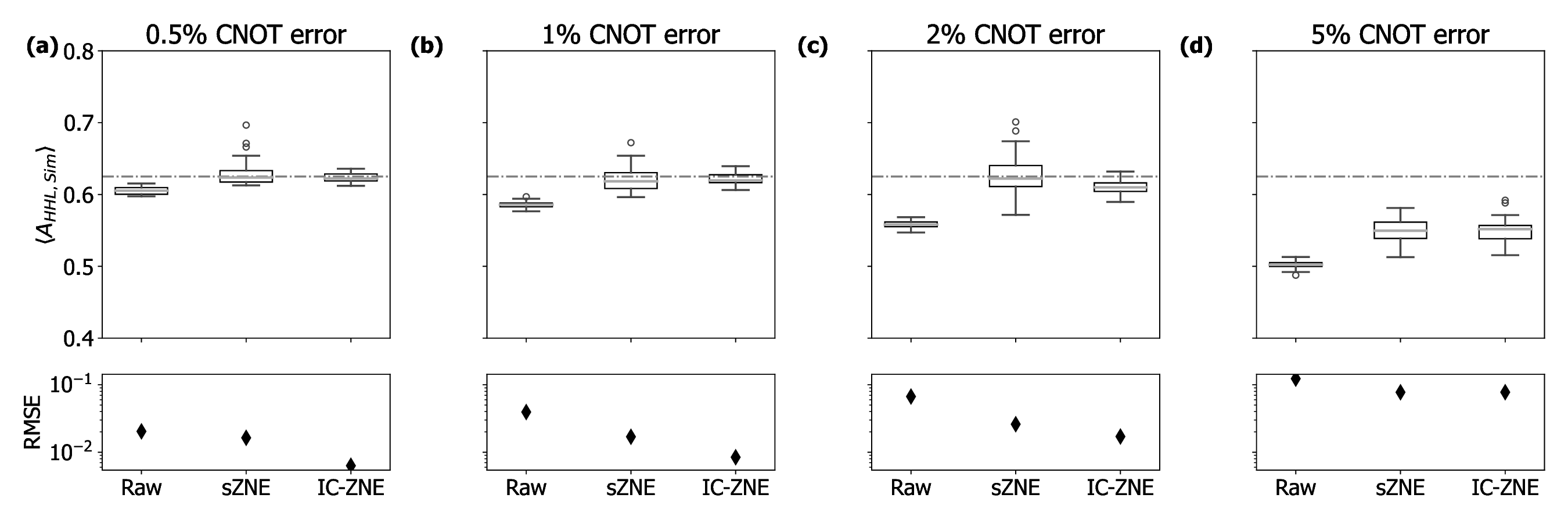}
    \caption{ Same as Fig.~\ref{fig:grover_simulation}, but for the HHL instead of the Grover circuit. IC-ZNE and sZNE show similar RMSE values. For CNOT error rates $\leq$~2\%, the distribution for sZNE is broader than for IC-ZNE.
    IC-ZNE performs at a minimum of 1.6 times (in terms of RMSE) better than sZNE in (a), (b) and (c).
    For higher error rates, the obtained extrapolated values significantly differ from the exact value (horizontal line). This indicates that, due to the larger number of CNOT gates as compared to the Grover circuit, the errors in
    (d) are too large for ZNE to work effectively.}
    \label{fig:hhl_simulation}
\end{figure*}

The results of $M=50$ simulation runs per error model are shown  in Fig.~\ref{fig:grover_simulation} (Grover) and Fig.~\ref{fig:hhl_simulation} (HHL).
In both cases, we perform sZNE and IC-ZNE. For comparison, the raw data obtained from the unscaled circuit without performing ZNE is also displayed. Note that randomized compiling is not employed in the simulations, since it has no impact on a depolarizing channel.
In each case, we obtain $M=50$ extrapolated values $\langle A_{\rm Grover}\rangle_i$ and $\langle A_{\text{HHL} }\rangle_i$ from the different simulation runs ($i=1,2,\dots,M)$. From these values, we generate a box plot showing the median value (central line) and the interquartile range (box). The whiskers extend to the highest and lowest values within 1.5 times the interquartile range. Any data points outside of these boundaries are considered outliers and are plotted separately as circles. The accompanying plots at the bottom display the root mean square deviation from the exact value:
\begin{equation}
    {\rm RMSE}=\sqrt{\frac{1}{M}\sum_{i=1}^M \Bigl(\langle A\rangle_i-\langle A\rangle_{\rm ideal}\Bigr)^2},
\end{equation}
where $A=A_{\rm Grover}$ or $A_{\rm HHL}$, respectively. 

From Figs.~\ref{fig:grover_simulation} and \ref{fig:hhl_simulation}, it is evident that, both for the Grover and the HHL circuit, and for all four error models introduced above (with different error rates of the CNOT and
single-qubit gates), IC-ZNE
demonstrates superior performance compared to sZNE, exhibiting the smallest RMSE in all  but one instance (corresponding to HHL with largest error rate, see below). 

For the Grover circuit (Fig.~\ref{fig:grover_simulation}),
we observe a slight bias of IC-ZNE towards larger values of $\langle A_{\rm Grover}\rangle$ for the two highest error rates (c) and (d). IC-ZNE performs at a minimum of 1.6 times
better (in terms of RMSE) than sZNE. In the best case, see Fig.~\ref{fig:grover_simulation}(b), IC-ZNE exhibits a 3.2-fold improvement over sZNE by comparison of the RMSE values.

For the HHL circuit,  Figs.~\ref{fig:hhl_simulation} (a), (b) and (c) demonstrate that sZNE exhibits a greater degree of variation than IC-ZNE. For the lowest error rate in (a), IC-ZNE performs 2.5 times better, because it has fewer variations and outliers. For CNOT error rates
reaching 5\%, see Fig.~\ref{fig:hhl_simulation} (d),
the extrapolated values of, both, sZNE and IC-ZNE deviate significantly from the exact value (horizontal line).
This indicates a fundamental limitation of ZNE techniques, which cannot be expected to yield accurate results if the level of noise in the original circuit is too high (due to the larger number of CNOT gates in comparison to the Grover circuit). 

\subsection{\label{sec:benchmark}Runs on quantum computing hardware and effects of twirling}

\begin{figure*}
    \includegraphics[width=\textwidth]{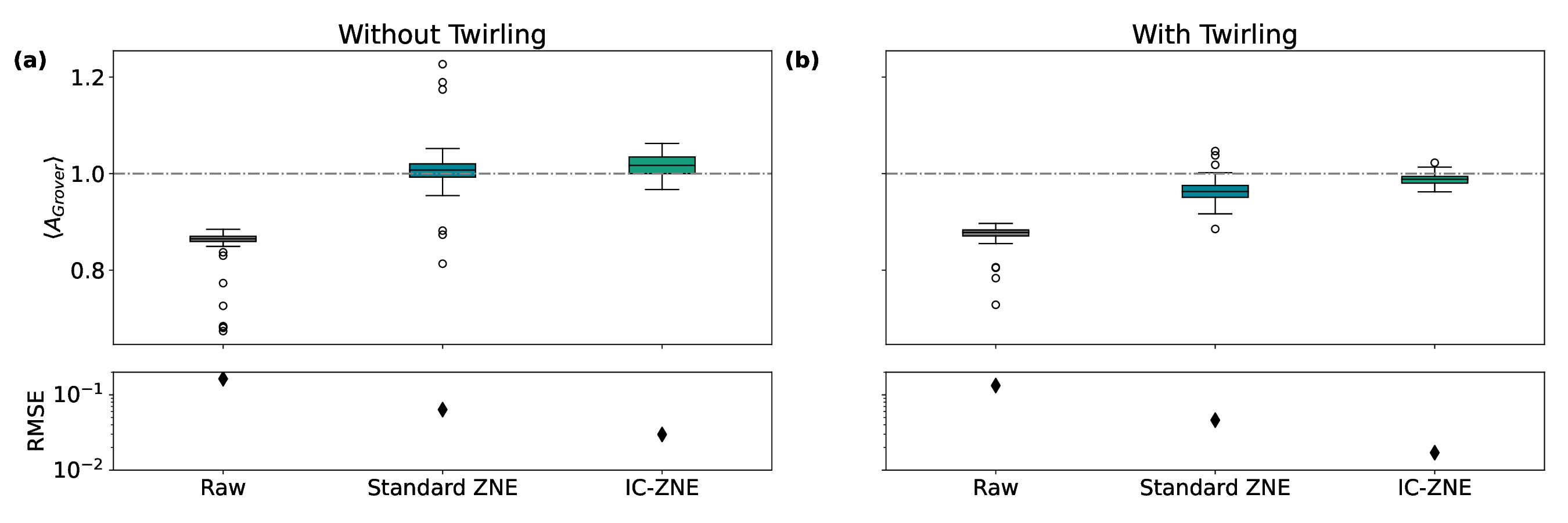}
    \caption{\label{fig:grover_backend}Comparison of inverted-circuit ZNE (IC-ZNE), standard ZNE (sZNE) and unmitigated raw data (Raw) for 50 runs of the Grover circuit on the IBM quantum system \textit{ibmq\_ehningen} without (a) and with (b) randomized compiling by Pauli twirling. Like in Fig.~\ref{fig:grover_simulation}, the obtained values of $\langle A_{\rm Grover}\rangle$ are shown in a box plot. The corresponding root-mean-square errors (RMSE) indicating the deviations from the exact value $\langle A_{\rm Grover}\rangle_{\rm ideal}=1$ (horizontal dashed line) are displayed in the accompanying plots below. The most accurate result -- with smallest RMSE and a smaller number of statistical outliers (open circles) -- is obtained using IC-ZNE with twirling.}
\end{figure*}

\begin{figure*}
    \includegraphics[width=\textwidth]{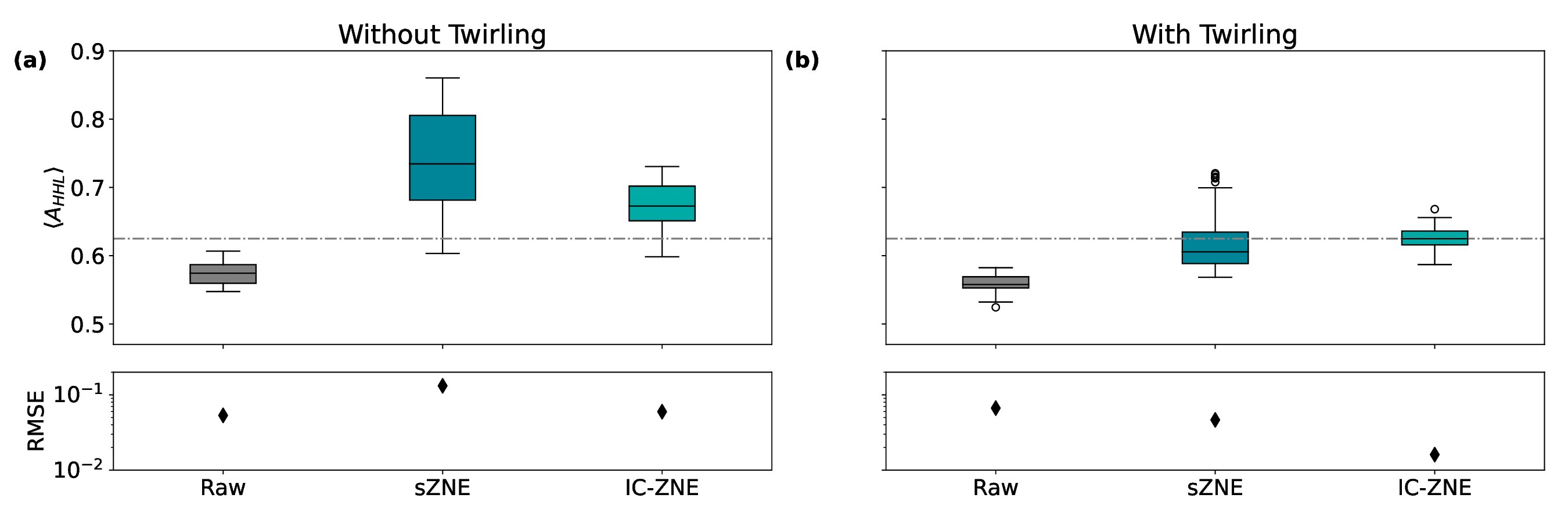}
    \caption{Same as Fig.~\ref{fig:grover_backend}, but for the HHL instead of Grover circuit.
    (a) Data gathered without twirling. The expectation values of both mitigation techniques show no improvement as compared to the unmitigated raw data. (b) With twirling, IC-ZNE improves the outcome, achieving a root-mean-square error three times smaller than sZNE.}
    \label{fig:hhl_backend}
\end{figure*}

The above results, which show
the superiority of IC-ZNE over sZNE,
but also the limitations in the case of excessive error rates, were obtained on simulators with depolarizing noise. On real quantum hardware, however, the errors have a more complex structure: e.g. correlations or crosstalk between neighboring qubits \cite{ketterer2023} or coherent errors can occur alongside incoherent ones. The technique of randomized compiling simplifies this structure by transforming the errors occurring in CNOT gates into an incoherent Pauli channel (see Sec.~\ref{sec:additional techniques}). Therefore, we now compare the results of 50 extrapolation runs obtained on real hardware (\textit{ibmq\_ehningen}) with or without randomized compiling. The fitting procedure without randomized compiling is the same as described in Sec.~\ref{sec:fitting_procedure}, but with 16 identical, instead of different randomly twirled circuits. 
Moreover, we apply readout error mitigation based on the method M3 \cite{Nation2021}, both for determining the expectation values $\langle A\rangle$ and the error strengths $\epsilon$. 

For the Grover circuit,
Fig. \ref{fig:grover_backend} depicts the raw data without (a) and with (b) twirling gates and the mitigated expectation values constructed by exponential extrapolation using sZNE or linear extrapolation using IC-ZNE, respectively. The boxplot shows that the distance between the first and the third quartile of $A_{Grover}$ (i.e., the size of the box) is higher for IC-ZNE in the case without twirling, but, in both cases (with and without twirling), there are fewer or no outliers (open circles) compared to raw data or sZNE. 
Moreover, the median expectation value obtained by IC-ZNE with twirling is closer to the exact value, resulting in a small root-mean-square error (RSME).  
As discussed in Sec.~\ref{sec:fitting_procedure}, this excellent performance of IC-ZNE, which provides an estimation essentially without bias, 
is due to the more accurate determination of the error strength, if the errors are not described only by depolarizing noise.

The mitigated expectation values shown for the sZNE, both, with and without twirling gates display better accuracy (i.e., smaller RMSE) than the raw data. Twirling results in a smaller number of outliers, but, somewhat surprisingly, in a stronger deviation of the median from the exact value such that, in total, the RMSE turns out to be almost identical with and without twirling.

In contrast, when using IC-ZNE, it is advantageous to utilize twirling, as it improves the mitigation result and accuracy compared to sZNE and raw data. This can also be seen, when examining the root-mean-square error (RMSE) in Fig. \ref{fig:grover_backend} (lower part), where it is evident that IC-ZNE outperforms sZNE.

Fig.~\ref{fig:hhl_backend} shows the results of 50 runs of the HHL circuit without (a) and with (b) twirling on the backend. As can be seen in Fig.~\ref{fig:hhl_backend}(a), the RMSE values for ``Raw''
and ``IC-ZNE'' are in the same order of magnitude. However, ``sZNE'' shows a wider range of values, resulting in a higher RMSE value.
Upon inspection of the median values (horizontal lines within the boxes), it is evident that both mitigation methods overestimate the expectation value without twirling.

If twirling is included, see Fig.~\ref{fig:hhl_backend}(b), the accuracy of IC-ZNE improves significantly, with perfect alignment between the median value of $\langle A_{\text{HHL} }\rangle$ and the exact result. Consequently, the root-mean-square error is three times lower for IC-ZNE than for sZNE.

Finally, we would like to comment on the additional resources required for IC-ZNE. As compared to standard ZNE, the execution time on the quantum hardware is approximately twice as large due to the additional circuits required for the measurement of error strengths. The fact that, due to addition of the inverse, these circuits are also twice as deep has little impact on the execution time, which, in present IBM devices, is mainly limited by the preparation of the initial state and hence approximately proportional to the total number of shots.

The experiments discussed in Secs.~\ref{sec:simulation} and \ref{sec:benchmark} 
were performed with 10~000 shots for the unmitigated raw data, 30~000 shots for standard ZNE and 60~000 shots for IC-ZNE, respectively. Alternatively, one could also perform standard ZNE and IC-ZNE with the same resources by either increasing the number of shots per circuit by a factor of two in case of standard ZNE or reducing it by a factor of two in case of IC-ZNE. Although this would sightly affect the variance of statistical fluctuations observed in Figs.~\ref{fig:grover_backend} and \ref{fig:hhl_backend}, we expect that the superiority of IC-ZNE compared to standard ZNE on the real hardware would  remain, since IC-ZNE (in combination with randomized compiling) provides estimations with essentially zero bias by avoiding systematic shifts due to a more accurate determination of noise scaling.

This expectation is confirmed by simulations shown in Appendix~\ref{sec:shots}, showing that the results of IC-ZNE do not strongly deteriorate when using half the number of shots.

\section{\label{sec:conclusion}Conclusions and Outlook}

This work introduced the method of Inverted-Circuit Zero-Noise Extrapolation (IC-ZNE), which is based on measuring the error strength using inverted circuits.
We presented its theoretical foundation, starting from a precise definition of the error strength $\epsilon$, from which we derived a relation between the error strength and the probability of measuring all qubits again in the initial state $|0\rangle$ after applying the circuit and its inverse. This relation is valid under the assumption of predominantly incoherent errors,  which can be justified by randomized compiling.
Our results in numerical simulations and on a quantum device clearly show the advantage of our method: it invariably delivers the smallest root-mean square deviation from the error-free value and provides estimations with essentially zero bias, unless the noise strength of the original circuit is too large.
The reason for the improved performance is the accurate determination of the scaling of the error strength, which deviates from the exponential scaling assumption made in standard ZNE (sZNE). 
Our method can be used for arbitrary circuits and does not require any knowledge of the underlying noise or complex error characterization.

In the future, it may be beneficial to explore alternative methods for scaling circuits, such as adaptive factors \cite{Krebs2022}. Furthermore, it remains to be explored how our method can be combined with other techniques, such as 
dual-state purification \cite{Cai_2021_puri, Huo_2022}, or
verifying whether the error strength of the scaled circuits obtained with probabilistic error amplification (PEA) scales as expected. Additionally, the IC-ZNE method can potentially be used to identify runs with high error rates, allowing for the selection of runs that cannot be mitigated.
\hfill\break
\hfill\break
The data that support the findings of this study are available from the corresponding author, upon reasonable request.

\begin{acknowledgments}
The authors acknowledge funding from the Ministry of Economic Affairs, Labor and Tourism Baden-W{\"u}rttemberg, under the project Sequoia - End to End in the frame of the Competence Center Quantum Computing Baden- W{\"u}rttemberg.
\end{acknowledgments}

\appendix

\section{\label{sec:level5}Adjoint noise channel}

Here, we justify the assumption that the channel ${\mathcal E}_{U^\dagger}^\dagger$ exhibits the same fidelity as ${\mathcal E}_{U}$, see Eq.~(\ref{eq:rhotilde_decomp}) in Sec.~\ref{sec:level4}.

Every circuit can be decomposed into single-qubit gates (denoted by $S$ in the following) and CNOT gates ($C$), for example:
\begin{equation}
U=S_3 C_2 S_2 C_1 S_1.
\end{equation}
Concerning $U^\dagger$, we use the fact that CNOT gates are self-adjoint:
\begin{equation}
U^\dagger=S_1^\dagger C_1 S_2^\dagger C_2 S_3^\dagger.
\end{equation}
In the following, we ignore the error of the single-qubit gates. If we apply randomized compiling, the errors of the CNOT gates are given by Pauli channels
(i.e. $C\to \Lambda_{\mathcal P} {\mathcal C}$):
\begin{equation}
{\mathcal E}_U = {\mathcal S}_3 \Lambda_{{\mathcal P}_2} {\mathcal C}_2 {\mathcal S}_2 \Lambda_{{\mathcal P}_1} {\mathcal C}_1 {\mathcal S}_1.
\end{equation}
Similarly for $U^\dagger$:
\begin{equation}
{\mathcal E}_{U^\dagger} = {\mathcal S}_1^\dagger\Lambda_{{\mathcal P}_1} {\mathcal C}_1 {\mathcal S}_2^\dagger\Lambda_{{\mathcal P}_2}{\mathcal C}_2{\mathcal S}_3^\dagger.
\label{eq:EU}
\end{equation}
Using the fact that Pauli channels are self-adjoint, the adjoint channel finally turns out as:
\begin{equation}
{\mathcal E}_{U^\dagger}^\dagger = {\mathcal S}_3 {\mathcal C}_2\Lambda_{{\mathcal P}_2} {\mathcal S}_2{\mathcal C}_1\Lambda_{{\mathcal P}_1}{\mathcal S}_1.
\end{equation}
It is almost the same as ${\mathcal E}_U$, see Eq.~(\ref{eq:EU}), except that the noise channels are applied before instead of after the ideal  CNOT operations. Let us consider a Pauli operator $P_i$ which does not commute with CNOT. In this case, the commutation leads to another Pauli operator $P_{\pi(i)}$, see Eq.~(\ref{eq:Pauli_CNOT_commutation}):
\begin{equation}
P_i C = C P_{\pi(i)}.
\end{equation}
Therefore, ${\mathcal E}_{U^\dagger}$ and ${\mathcal E}_{U^\dagger}^\dagger$ may in principle differ for Pauli channels where
$p_i\neq p_{\pi(i)}$ for at least one $i$. However, we expect the effect on the fidelity to be small, since the latter mainly originates from the Pauli terms proportional to the identity operator $P_1$ (no error), which commutes with CNOT. 

\section{\label{sec:transpiled_circuits}Transpiled circuits}
\begin{figure*}[ht]
    \includegraphics[width=\textwidth]{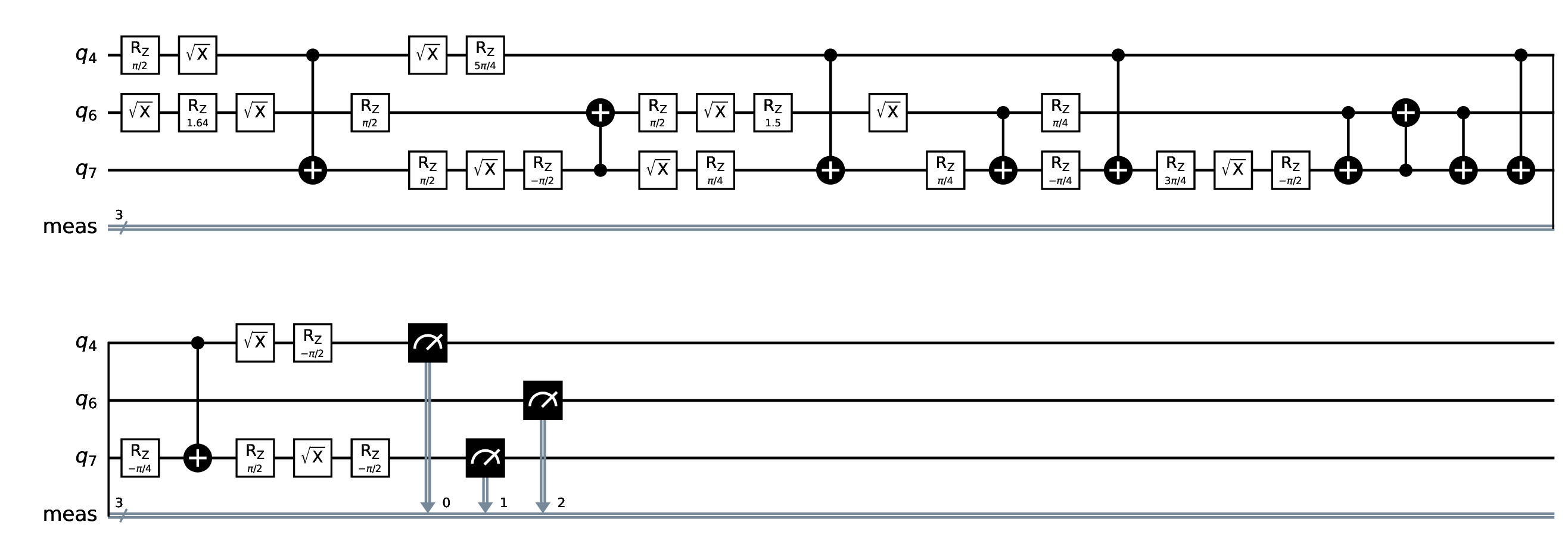}
    \caption{Representation of the Grover circuit which is used in Sec.~\ref{sec:results} to test the performance of ZNE. The circuit has been designed to conform to IBM's quantum device architecture through transpilation. This requires the use of one SWAP gate, which consists of three CNOT gates.}
    \label{fig:qc_t_grover}
\end{figure*}

\begin{figure*}[ht]
    \includegraphics[width=\textwidth]{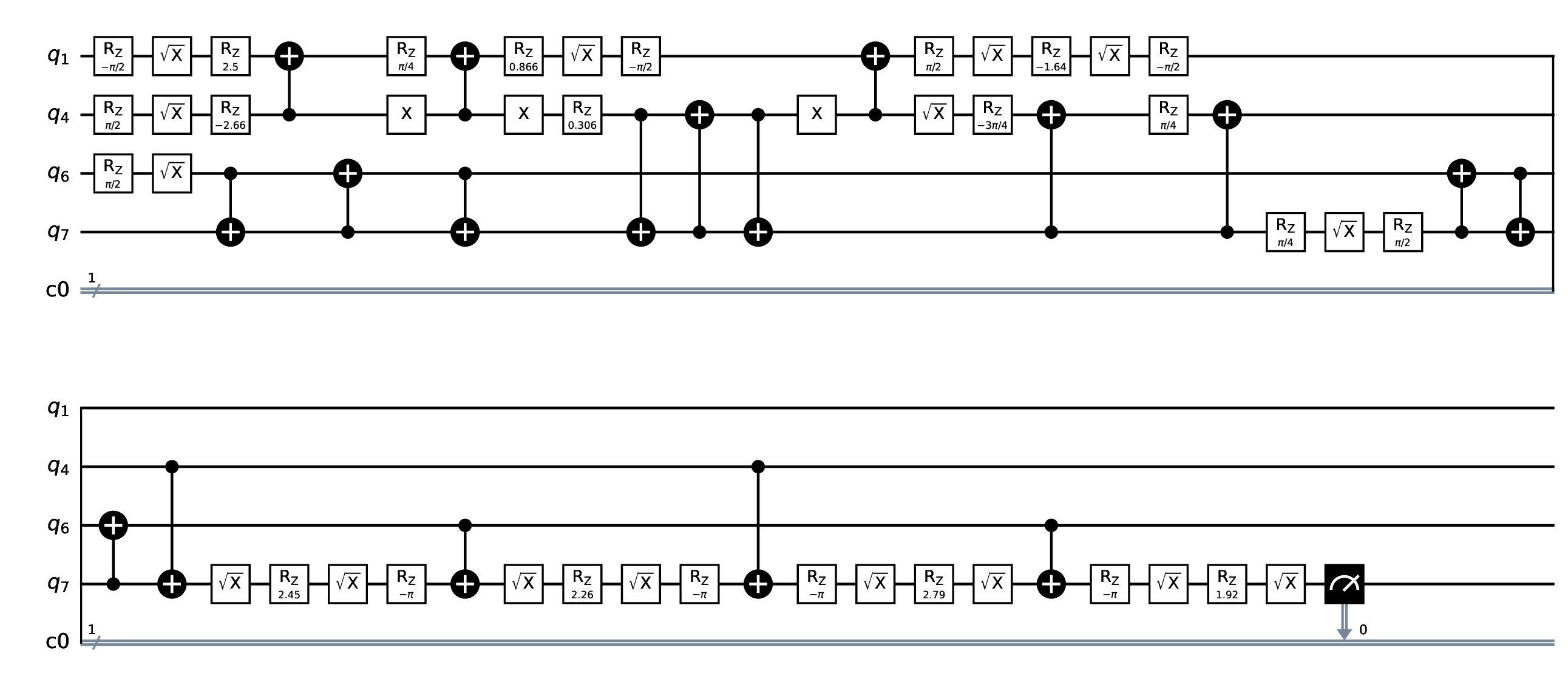}
    \caption{Representation of the HHL circuit which is used in Sec.~\ref{sec:results} to test the performance of ZNE. Note that only the last qubit ($q_3$) is measured in case of HHL. The circuit has been designed to conform to IBM's quantum device architecture through transpilation. This requires the use of three SWAP gates, each of which consists of three CNOT gates.}
    \label{fig:qc_t_hhl}
\end{figure*}

In Figs.~\ref{fig:qc_t_grover} and \ref{fig:qc_t_hhl}, we illustrate the quantum circuits utilized in this paper, specifically tailored to conform to IBM's quantum device architecture through transpilation. Both circuit diagrams incorporate at least one SWAP gate, necessitated by the constraints of the device architecture, thereby augmenting the total count of CNOT gates. The Grover circuit showcased in Fig.~\ref{fig:qc_t_grover} requires only one SWAP gate, resulting in a cumulative total of 10 CNOT gates. In contrast the HHL circuit depicted in Fig.~\ref{fig:qc_t_hhl} exhibits a total of 18 CNOT gates, of which 9 are attributed solely to the inclusion of SWAP gates.

\section{\label{sec:shots}Simulation of IC-ZNE with reduced number of shots}

\begin{figure*}[ht]
    \includegraphics[width=\textwidth]{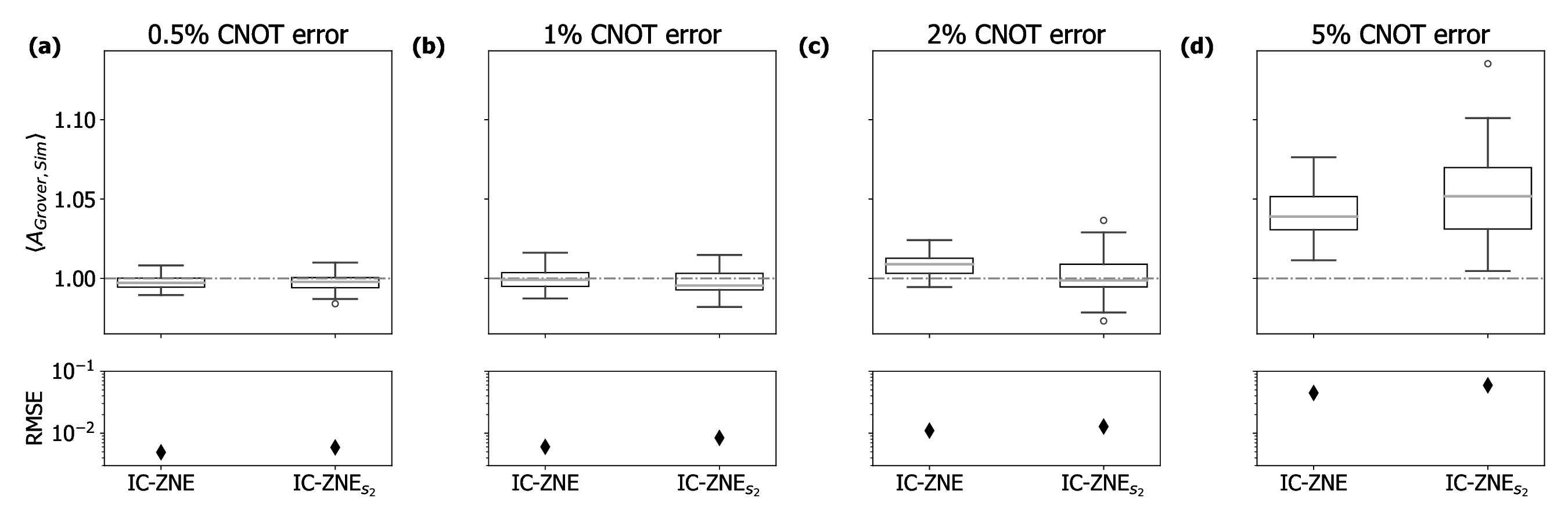}
    \caption{
    Results of Fig.~\ref{fig:grover_simulation} are compared with results obtained when using 30.000 (IC-ZNE$_{s_2}$) instead of 60.000 shots (IC-ZNE).
     \label{fig:grover_lessShots}
     }
\end{figure*}

In Fig.~\ref{fig:grover_lessShots}, we compare the simulation results of IC-ZNE obtained with 30.000 and 60.000 shots in total. The quality of the result, measured in terms of the root mean-square error (RSME), is almost the same in both cases.

\section{\label{sec:supplement}Device properties}

\begin{figure}
    \includegraphics[width=0.5\textwidth]{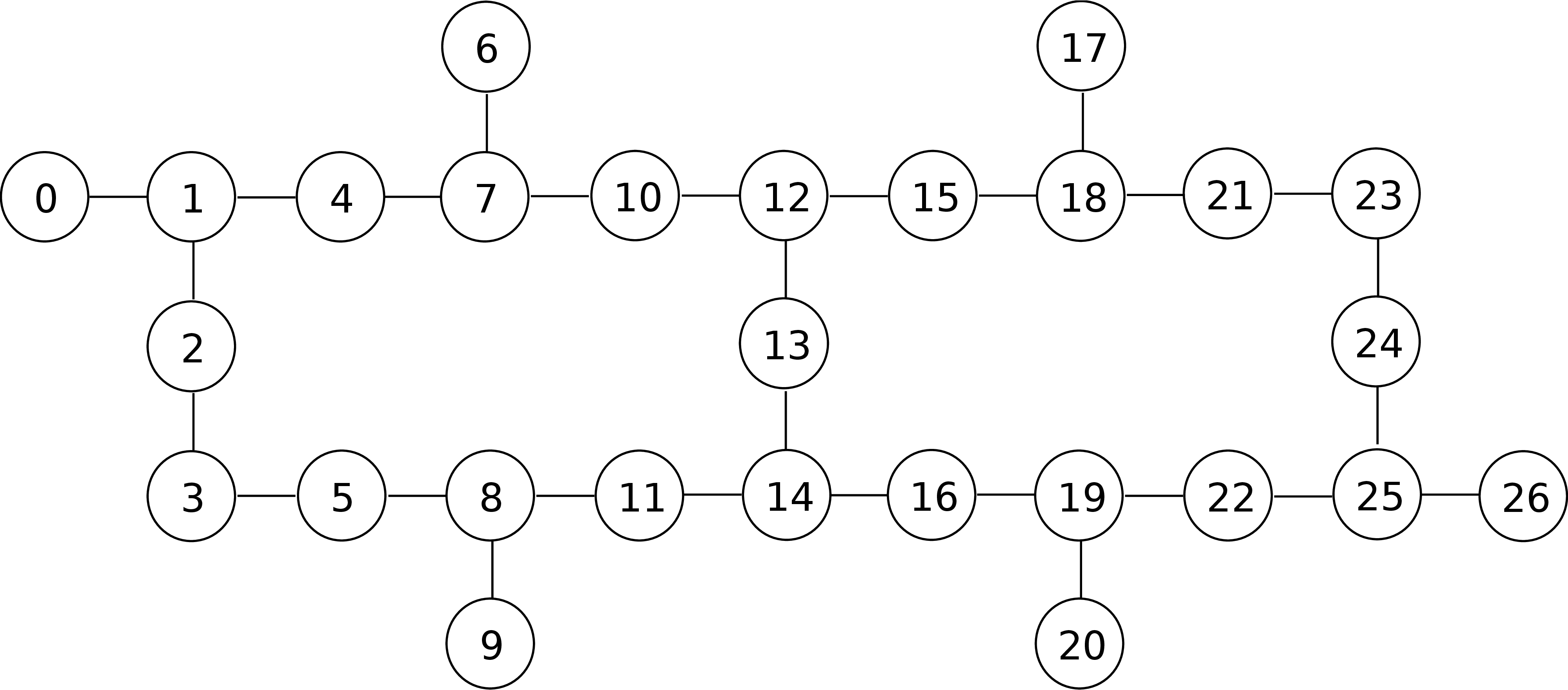}
    \caption{Connectivity (black lines) between 27 qubits (black circles) of IBM's Ehningen device.}
    \label{fig:layout}
\end{figure}

\begin{figure*}[ht]
    \includegraphics[width=\textwidth]{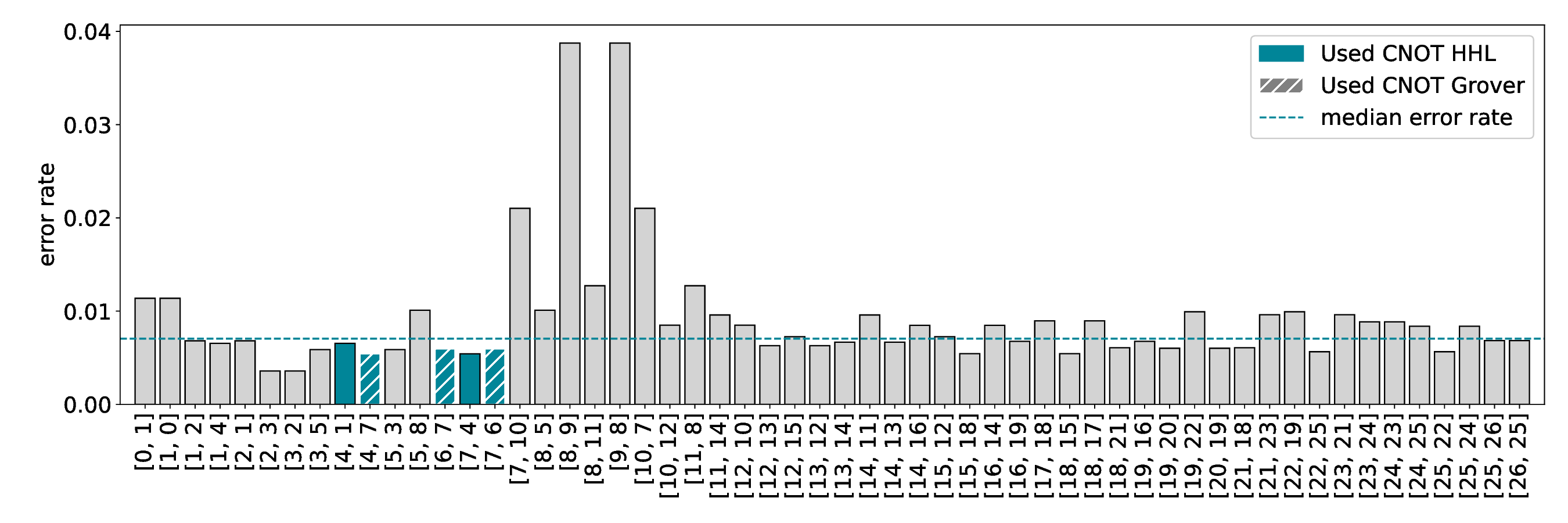}
    \caption{Histogram of the error rates for each CNOT gate on IBM's cloud device \textit{imbq\_ehningen} from August 18th, 2023. The dashed green line represents the median error rate of the device. The green bars represent the CNOT gates used for runs with the HHL circuit ([4, 1],[6, 7], [7, 6],[4, 7],[7, 4],[4, 1],[6, 7]), while the hatched bars represent those used for runs with the Grover circuit ([4, 7],[6, 7],[7, 6]). The reported error rates for the CNOT gates used are lower than the median error rate and are below 1\%.}
    \label{fig:error_rates}
\end{figure*}

This section presents the detailed specifications of the data obtained from IBM's quantum device, as described in the main text. Fig.~\ref{fig:layout} illustrates the connectivity of the device \textit{ibmq\_ehningen}, which hosts 27 qubits in a heavy hexagonal lattice.

The error rates of all CNOT gates of \textit{ibmq\_ehningen} (from August 18th, 2023) as reported by IBM are displayed in Fig.~\ref{fig:error_rates}. We observe variations in error rates across different CNOT gates, which can significantly impact the success probability of quantum algorithms. The median error rate, indicated by the dashed green line, is 0.705\%. The CNOT gates used for runs involving the HHL circuit are represented by the green bars ([4, 1],[6, 7], [7, 6],[4, 7],[7, 4],[4, 1],[6, 7]), while the hatched bars represent those employed for runs with the Grover circuit ([4, 7],[6, 7],[7, 6]). The precise error rates for each qubit pair is presented in Table~\ref{table:cxgate_properties}.

Note that these error rates only give a partial description of the noise occurring on the real device (e.g., they do not include crosstalk effects \cite{ketterer2023}). Therefore, the total error strengths measured using the method of inverted circuits ($\epsilon_0\simeq 0.15$ for the Grover circuit and $\epsilon_0\simeq 0.4$ for the HHL circuit, see the green diamonds in Figs.~\ref{fig:grover_std_vs_iczne} and \ref{fig:hhl_std_vs_iczne}) are typically larger than expected from the error rates per CNOT gate displayed in Fig.~\ref{fig:error_rates}.  

The single-qubit device properties of \textit{ibmq\_ehningen} as reported by IBM are presented in Table~\ref{table:backend_properties}.

\begin{table}[h!]
    \centering
    \caption{Table of the gate error of the available two-qubit gate pairs for \textit{ibmq\_ehningen} (as of August 18th, 2023).}
    \label{table:cxgate_properties}
    \begin{tabular*}{0.5 \textwidth}{@{\extracolsep{\fill}}cccc}
        CX Gate Pair & Gate error & CX Gate Pair & Gate error \\ \hline
        25\_22 & 0.00565 & 10\_12 & 0.00849 \\ 
        22\_25 & 0.00565 & 12\_10 & 0.00849 \\ 
        20\_19 & 0.00602 & 7\_10 & 0.02103 \\
        19\_20 & 0.00602 & 10\_7 & 0.02103 \\ 
        14\_16 & 0.00847 & 19\_22 & 0.00993 \\ 
        16\_14 & 0.00847 & 22\_19 & 0.00993 \\ 
        18\_17 & 0.00896 & 21\_23 & 0.00961 \\ 
        17\_18 & 0.00896 & 23\_21 & 0.00961 \\ 
        14\_11 & 0.00959 & 13\_12 & 0.00629 \\ 
        11\_14 & 0.00959 & 12\_13 & 0.00629 \\ 
        12\_15 & 0.00726 & 26\_25 & 0.00684 \\ 
        15\_12 & 0.00726 & 25\_26 & 0.00684 \\
        8\_5 & 0.0101 & 3\_2 & 0.00358 \\ 
        5\_8 & 0.0101 & 2\_3 & 0.00358 \\ 
        1\_0 & 0.01138 & 7\_6 & 0.00595 \\ 
        0\_1 & 0.01138 & 6\_7 & 0.00595 \\ 
        1\_2 & 0.00681 & 16\_19 & 0.00676 \\
        2\_1 & 0.00681 & 19\_16 & 0.00676 \\ 
        8\_11 & 0.01272 & 18\_21 & 0.00607 \\
        11\_8 & 0.01272 & 21\_18 & 0.00607 \\
        24\_23 & 0.00885 & 4\_7 & 0.00542 \\ 
        23\_24 & 0.00885 & 7\_4 & 0.00542 \\ 
        4\_1 & 0.00655 & 14\_13 & 0.00667 \\ 
        1\_4 & 0.00655 & 13\_14 & 0.00667 \\ 
        8\_9 & 0.03875 & 24\_25 & 0.00839 \\ 
        9\_8 & 0.03875 & 25\_24 & 0.00839 \\ 
        18\_15 & 0.00544 & 3\_5 & 0.00588 \\ 
        15\_18 & 0.00544 & 5\_3 & 0.00588 \\ 
    \end{tabular*}
\end{table}

\begin{table*}[ht]
    \centering
    \caption{Table of the single-qubit device properties of \textit{ibmq\_ehningen} (as of August 18th, 2023). The table displays the frequencies, T1 and T2, gate errors, and readout errors for the various qubits, as reported by IBM.}
    \label{table:backend_properties}
    \begin{tabular*}{0.9 \textwidth}{@{\extracolsep{\fill}}cccccccc}
        Qubit & Frequency / GHz & T1 /µs & T2 /µs & RZ error & SX error & X error& Readout error \\ \hline
        Q0 & 4.961 & 142.59275 & 95.47855 & 0 & 0.00018 & 0.00018 & 0.011 \\
        Q1 & 5.18191 & 181.00386 & 70.72969 & 0 & 0.00021 & 0.00021 & 0.0096 \\
        Q2 & 5.12694 & 95.06903 & 9.06901 & 0 & 0.0003 & 0.0003 & 0.009 \\
        Q3 & 5.26815 & 99.89562 & 28.579 & 0 & 0.00017 & 0.00017 & 0.0229 \\
        Q4 & 5.05357 & 159.58091 & 70.14323 & 0 & 0.00068 & 0.00068 & 0.0123 \\
        Q5 & 5.07116 & 104.48761 & 8.15402 & 0 & 0.00032 & 0.00032 & 0.024 \\ 
        Q6 & 4.89006 & 143.66941 & 122.05597 & 0 & 0.00035 & 0.00035 & 0.014 \\
        Q7 & 4.97776 & 123.99653 & 113.28929 & 0 & 0.00021 & 0.00021 & 0.0076 \\
        Q8 & 5.17419 & 84.87928 & 69.68872 & 0 & 0.00104 & 0.00104 & 0.0245 \\ 
        Q9 & 4.9925 & 112.5169 & 62.87694 & 0 & 0.00573 & 0.00573 & 0.0175 \\ 
        Q10 & 4.83511 & 191.01486 & 162.14782 & 0 & 0.00039 & 0.00039 & 0.0092 \\
        Q11 & 5.11944 & 92.86756 & 127.55782 & 0 & 0.00027 & 0.00027 & 0.017 \\
        Q12 & 4.72549 & 169.03301 & 238.81786 & 0 & 0.00015 & 0.00015 & 0.0117 \\
        Q13 & 4.92598 & 155.88103 & 225.73924 & 0 & 0.00022 & 0.00022 & 0.008 \\ 
        Q14 & 5.17671 & 94.24501 & 258.67336 & 0 & 0.00033 & 0.00033 & 0.0077 \\ 
        Q15 & 4.89299 & 93.54811 & 176.56123 & 0 & 0.00019 & 0.00019 & 0.0105 \\ 
        Q16 & 5.02214 & 194.67907 & 191.30807 & 0 & 0.00028 & 0.00028 & 0.0078 \\ 
        Q17 & 5.13566 & 146.76432 & 22.47776 & 0 & 0.00052 & 0.00052 & 0.0073 \\ 
        Q18 & 4.99642 & 130.38322 & 228.58117 & 0 & 0.00024 & 0.00024 & 0.0128 \\
        Q19 & 4.7841 & 172.28179 & 77.40924 & 0 & 0.00032 & 0.00032 & 0.0137 \\ 
        Q20 & 5.04235 & 204.49692 & 220.82418 & 0 & 0.00053 & 0.00053 & 0.025 \\ 
        Q21 & 4.93974 & 126.28787 & 202.81237 & 0 & 0.00026 & 0.00026 & 0.0076 \\
        Q22 & 4.72513 & 58.42632 & 34.32104 & 0 & 0.00025 & 0.00025 & 0.0131 \\
        Q23 & 4.80479 & 190.68715 & 238.40096 & 0 & 0.00043 & 0.00043 & 0.0084 \\
        Q24 & 5.07449 & 233.74912 & 337.44479 & 0 & 0.00015 & 0.00015 & 0.0075 \\
        Q25 & 4.95019 & 198.39525 & 439.17036 & 0 & 0.00018 & 0.00018 & 0.0076 \\ 
        Q26 & 5.15132 & 190.73366 & 24.68715 & 0 & 0.00015 & 0.00015 & 0.0079 \\
    \end{tabular*}
\end{table*}

\clearpage
\bibliography{00_main}

\end{document}